\RequirePackage{fix-cm}
\documentclass[smallextended]{svjour3}       
\smartqed  
\usepackage{graphicx}
\usepackage[T1]{fontenc}
\usepackage{lmodern}
\usepackage{mathptmx}
\usepackage[scaled=0.9]{helvet}
\usepackage{courier}
\usepackage[T1]{fontenc}
\usepackage{mathrsfs}
\usepackage{subfigure}
\usepackage{multirow}
\usepackage{url,amsfonts,epsfig}
\usepackage{fancyhdr}
\usepackage{color}
\usepackage{subfigure}
\usepackage{subfigmat}
\usepackage{anyfontsize}
\usepackage{ulem}
\usepackage{pstricks,pst-sigsys,pst-eps}
\usepackage{multirow}
\usepackage{booktabs}
\usepackage{fancybox}
\usepackage{mathptmx}
\usepackage{anyfontsize}
\usepackage{t1enc}
\usepackage{amssymb,latexsym}
\usepackage{amsmath,amstext,amssymb,latexsym, psfrag}
\usepackage{graphicx,colortbl, epsfig, subfigure}
\usepackage{threeparttable}
\usepackage{float}

\newcommand{\bn}{{\mathbf {n}}}

\renewcommand{\u}{\boldsymbol{u}}
\newcommand{\U}{\boldsymbol{U}}

 \usepackage{epstopdf}
\definecolor{airforceblue}{rgb}{0.36, 0.54, 0.66} 

\begin{document}

\title{Assessment of a non-conservative four-equation multiphase system with phase transition 
}


\author{Paola Bacigaluppi  \and
        Julien Carlier  
        \and Marica Pelanti \and
        Pietro Marco Congedo \and R\'emi Abgrall 
}


\institute{P.  Bacigaluppi\at
Institute of Mathematics, University of Zurich, Switzerland \\
             \textit{Present address:} Department of Aerospace Science and Technology, Politecnico di Milano, Italy\\
              \email{paola.bacigaluppi@polimi.it}           
              \and
           J.  Carlier, M.  Pelanti \at
              ENSTA Paris, Institut Polytechnique de Paris, France
           \and
           P.M.  Congedo \at
              Inria, Centre des Math\'ematiques Appliqu\'ees, Ecole Polytechnique, France
              \and 
              R. Abgrall \at
              Institute of Mathematics, University of Zurich, Switzerland \\
}

\date{Received: date / Accepted: date}

\maketitle
\begin{abstract}
This work focuses on the formulation of a four-equation model for simulating unsteady two-phase mixtures with phase transition and strong discontinuities. The main assumption consists in a homogeneous temperature, pressure and velocity fields between the two phases. 

Specifically, we present the extension of a residual distribution scheme to solve a four-equation two-phase system with phase transition written in a non-conservative form, i.e. in terms of internal energy instead of the classical total energy approach. 
This non-conservative formulation allows avoiding the classical oscillations obtained by many approaches, that might appear for the pressure profile across contact discontinuities.
The proposed method relies on a Finite Element based Residual Distribution scheme which is designed for an explicit second-order time stepping. 

We test the non-conservative Residual Distribution scheme on several benchmark problems and assess the results via a cross-validation with the approximated solution obtained via a conservative approach, based on a HLLC scheme. 
Furthermore, we check both methods for mesh convergence and show the effective robustness on very severe test cases, that involve both problems with and without phase transition.
\keywords{Non-conservative formulation \and hyperbolic problems \and compressible flows \and multiphase flows \and phase transition \and Residual Distribution scheme \and Godunov method \and diffuse interface model}

\end{abstract}

\section{Introduction}

In many different engineering applications, as steam turbines, turbo-pumps or trilateral flash cycles, and engines or wings, we encounter multiphase flows. These flows are complicated to characterize and sometimes show a very peculiar behaviour as a phase transition.
Experimental approaches are crucial to detect different regimes according to the operating conditions but are highly expensive. Therefore, numerical simulations are massively used to predict the physical behaviour that might be encountered.

Among the mathematical models used in the multiphase community, some are based on the reinterpretation of the Euler equations with some additional source term. In particular, one may refer to the Baer-Nunziato seven-equation model, first presented in \cite{Baer1986}, from which several methods, with higher or lower complexity according to the physics aimed to be represented, have been derived (see \cite{Lund2012}). 
The present work relies on diffuse interface models. Specifically, it considers a reduced version of the Baer-Nunziato model, namely a four-equation model, where the reduction from the original seven-equation model comes from the assumption of considering that there is a very large interface between two different phases, such that the velocity, pressure and temperature of one phase coincide with those of the other media. 
This assumption is used, for example, in case the study concerns fog, which can be seen as a stratified two-phase flow, or cavitation, which is a highly violent implosive/explosive process.
This model features a thermo-chemical non-equilibrium, i.e. the involved flows undergo a mass transition. 
Moreover, let us note that one should not confuse the model used in this paper with a similar model, which was introduced in \cite{downar1996non} and is the so-called Homogeneous Relaxation  Model (HRM).

For solving the proposed four-equation system, we propose here a non-conservative formulation of the Residual Distributions Finite Element-type technique (RD) and cross-validate it with a more standard HLLC Finite Volume method  tailored for the conservative formulation of the considered model.
The proposed scheme aims at merging together different approximating strategies presented in previous papers, as \cite{AbgrallFVCA8,BacigaluppiNICFD2016,AbgrallCAF2017,BacigaluppiMood2018}. Indeed, the discretization is based on a second-order predictor-correction method as presented in \cite{Ricchiuto2010,AbgrallFVCA8,BacigaluppiNICFD2016,AbgrallHO2018}.
This extension has two objectives.
First, following the work of Abgrall et al. \cite{AbgrallCAF2017}, the considered system is rewritten in terms of a non-conservative formulation of the internal energy equation which is approximated in such a way, that the conservation is guaranteed.
The idea behind the choice of a non-conservative formulation, as seen in \cite{AbgrallCAF2017}, is due to the many advantages, especially for engineering applications, which are represented by the possibility to work directly with quantities, such as pressures or internal energies instead of having them derived from the total energy. 
Indeed, when coming, for example, across contact discontinuities, the velocity and the pressure do not change, while the density does. Deriving the internal energy or pressures from the total energy, may, therefore, not be completely accurate from a numerical point of view. Further, another relevant advantage is represented by the possibility of dealing easily with non-linear equations of state, as for example the family of Mie-Gr\"uneisen equations of state,  although in the present work a simple stiffened gas equation of state is used. To ensure a good approximation throughout the discontinuities, the so-called Multi-dimensional Optimal Order Detection (MOOD) limiting of \cite{BacigaluppiMood2018} has been applied, in order to switch from a second-order of accuracy to a first-order in case of numerical oscillations.

Secondly, we search to construct high-order residual distribution schemes for multiphase flow in multidimensional problems. In this perspective, using unstructured meshes may be of great interest in terms of geometrical flexibility and may permit a massive parallelization as they are at compact support. Consequently, we want to construct robust and efficient schemes for unstructured meshes.

In the following, we first introduce in Section \ref{sec:model} the mathematical model and recall the non-conservative formulation used to approximate the hyperbolic part of the considered system of equations. We then describe the Residual Distribution scheme and the treatment of the source terms accounting for the mass transition in Section \ref{PSIgalerkin}. Finally, in Section \ref{sec:cons} we briefly mention the Harten-Lax-van Leer-Contact (HLLC) approximate Riemann solver implemented for solving the conservative formulation of the four-equation model. In Section \ref{sec:results}, several significant benchmark problems cross-validate the chosen mathematical model and show the numerical validity of the non-conservative formulation of the system discretized via the Finite Element based Residual Distribution scheme. Finally, in Section \ref{sec:conclu}, we draw some conclusions and perspectives.

\section{The four-equation two-phase flow model}\label{sec:model}

\begin{figure}
\centering
\includegraphics[scale=0.22]{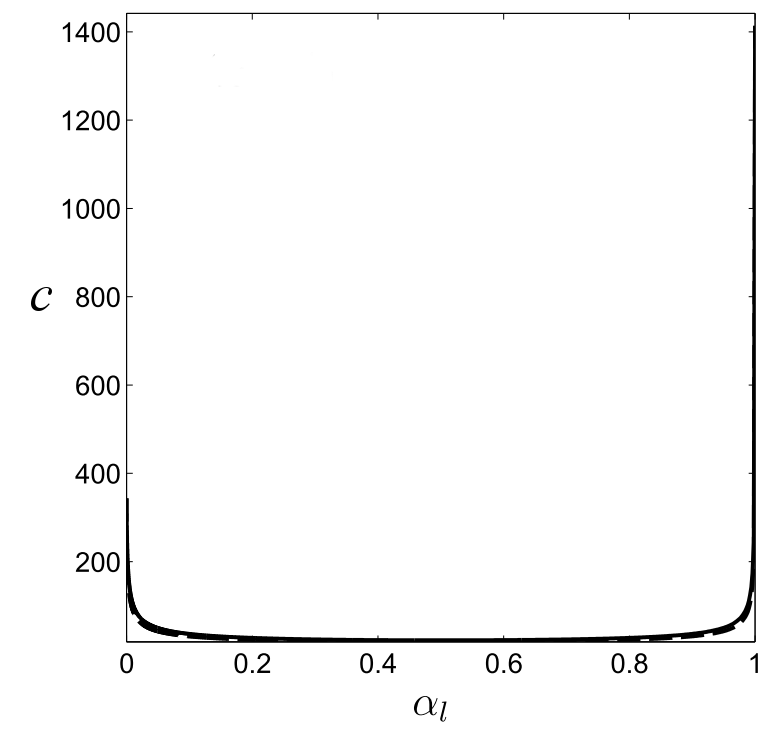}
\includegraphics[scale=0.22]{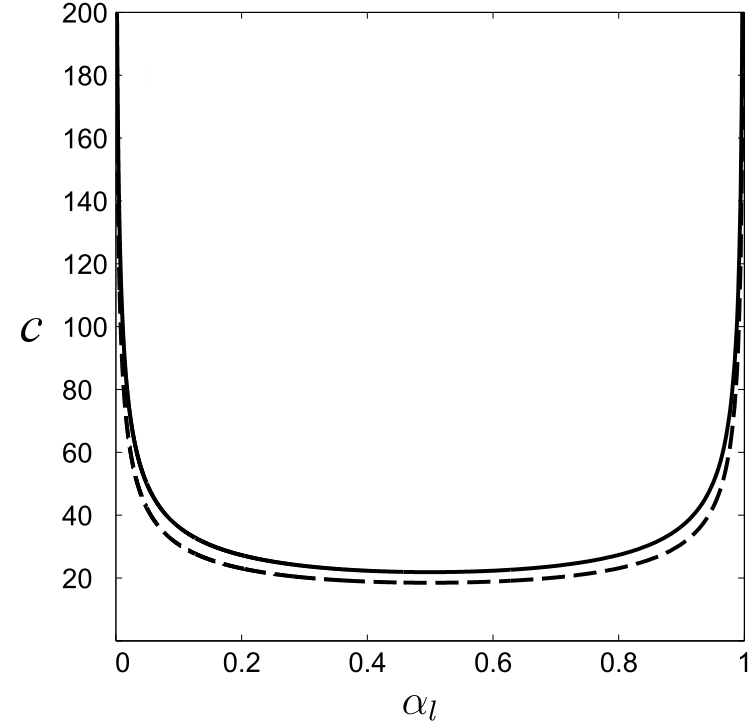}
\caption{Comparison between \eqref{Wood_formula} (continuous line) and \eqref{sound_4eqsn} (dashed line), for $\rho_{air}=1~kg/m^3$, $c_{air}=374~m/s$, $\rho_{water}=1000~kg/m^3$ and $c_{water}=1568~m/s$ and the phase values according to Table \ref{cbs_eos}. The right picture is a zoom of the left one.}
\label{sound_compare}
\end{figure}

We consider a four equation model that describes a two-phase flow in kinetic,
mechanical and thermal equilibrium \cite{LundAursand2012,le2014towards}.
This four-equation model can be obtained from the seven-equation Baer--Nunziato two-phase model \cite{Baer1986} in the limit of velocity,  pressure and temperature equilibrium.
It consists of one mass equation for each phase $k$ and momentum and energy equations for the mixture:
\begin{equation}
\begin{cases}
\partial_t (\alpha_1 \rho_1)+ \text{div}\;(\alpha_1 \rho_1 \; \textbf{u})=\Gamma,\\[0.5em]
\partial_t (\alpha_2 \rho_2)+ \text{div}\;(\alpha_2 \rho_2 \;\textbf{u})=-\Gamma,\\[0.5em]
\partial_t ( \rho \;\textbf{u})+\text{div}\;( \rho\;\textbf{ u} \otimes \textbf{u} +P \,Id)=0,\\[0.5em]
\partial_t E+ \text{div}\; \left(E+P\right) \textbf{u}=0.\\
\end{cases} 
\label{4eqsmodel_mass_c}
\end{equation}
Here  $\rho_k$ and $\alpha_k$ represent for each phase the density and the volume fraction, respectively.  The liquid (gaseous) phase is denoted by the subscript $l$ or $1$ ($g$ or $2$).
Moreover $\textbf{u}$ denotes the velocity, $\rho$ the mixture density, $P$ the pressure,
and $E$ the mixture total energy per unit volume, which is related to the mixture internal energy per unit volume
$e$ by
\begin{equation}
E=e+\frac{1}{2}\rho\textbf{u}^2,
\end{equation}
and $e=\sum_k \alpha_k e_k$, where $e_k$ is the internal energy of phase $k$.
 Note that the volume fraction $\alpha_k$  fulfills 
\begin{equation}
\sum_k \alpha_k=1.
\end{equation}
Here we rewrite the system (\ref{4eqsmodel_mass_c}) by replacing the conservative 
equation for the mixture total energy $E$ by a non-conservative equation
for the mixture internal energy $e$:
\begin{equation}
\begin{cases}
\partial_t (\alpha_1 \rho_1)+ \text{div}\;(\alpha_1 \rho_1 \; \textbf{u})=\Gamma,\\[0.5em]
\partial_t (\alpha_2 \rho_2)+ \text{div}\;(\alpha_2 \rho_2 \;\textbf{u})=-\Gamma,\\[0.5em]
\partial_t ( \rho \;\textbf{u})+\text{div}\;( \rho\;\textbf{ u} \otimes \textbf{u} +P \,Id)=0,\\[0.5em]
\partial_t e+ \textbf{u} \cdot \nabla e  + (e+P)\;\text{div}\; \textbf{u}=0.\\
\end{cases} 
\label{4eqsmodel_mass_nc}
\end{equation}
This four-equation model is hyperbolic and the associated speed of sound 
$c$ is defined by \cite{flatten-lund:rel}:
\begin{equation}
\label{sound_4eqsn}
\frac{1}{c^2} = \frac{1}{c_{Wood}^2} + \frac{\rho T C_{p1}C_{p2}}{C_{p1}+C_{p2}} 
\left(\frac{G_2}{\rho_2 c_2^2} -\frac{G_1}{\rho_1 c_1^2} \right)^2,
\end{equation}
where  $C_{pk}=\alpha_k \rho_k c_{pk}$, with $c_{pk}$ denoting the
specific heat capacity at constant pressure of phase $k$, and 
$G_k  = \left(\frac{\partial P_k}{\partial e_k}\right)_{\rho_k}$
(Gr\"uneisen coefficient). Above $T$ denotes the equilibrium temperature, $c_k$ denotes the speed of sound of phase $k$, and $c_{Wood}$ the Wood's speed of sound
\begin{equation}
\label{Wood_formula}
c_{Wood} = \left( \rho \sum_{k=1}^2 \frac{\alpha_k}{\rho_k c_k^2}\right)^{-\frac{1}{2}},
\end{equation}
 which corresponds to the
speed of sound of a flow in mechanical equilibrium
(but not thermal equilibrium). Note that the subcharacteristic condition
$c \leq c_{Wood}$ holds. This is also illustrated in Figure \ref{sound_compare}. 

The mass transfer term $\Gamma$ is equal to $\Gamma=\theta(g_l-g_g)$ with $g_l$, $g_v$ 
denoting the chemical potentials (Gibbs free energy) of the gaseous and liquid phases, while $\theta$ represents the relaxation parameter and it expresses  the inverse of the relaxation time of the process at which the thermodynamic equilibrium is reached.

We remark  that the four-equation model above is not equivalent to the 
 Homogeneous Relaxation Model (HRM) of \cite{downar1996non}.
 In fact the latter does not assume temperature equilibrium, so the thermodynamic closure of the two models is different.


The thermodynamic closure for \eqref{4eqsmodel_mass_c}-\eqref{4eqsmodel_mass_nc} reads
\begin{equation}\label{closure_mf4}
\begin{cases}
T_l=T_g=T\\[0.5em]
e=\alpha_l e_l+\alpha_g e_g\\[0.5em]
P_l=P_g=P\\[0.5em]
\alpha_l+\alpha_g=1.
\end{cases}
\end{equation}
Here $T_k$ represents the temperature for each phase $k$.
In the following, we consider the stiffened equations of state (EOS) for each phase $k$, that is 
characterized by the following relations:
\begin{equation}\label{thermo_4mf}
\begin{cases}
P_k(\rho_k,e_k)=(\gamma_k-1)(e_k-\rho_k\; q_k)-\gamma_k\; P_{\infty,k}\\[0.5em]
c_k(P_k,\rho_k)=\sqrt{\gamma_k \frac{P_k+P_{\infty,k}}{\rho_k}}\\[0.5em]
T_k(P_k,\rho_k)=\frac{P_k+P_{\infty,k}}{ \rho_k c_{v_{k}}\; (\gamma_k -1)}\\[0.5em]
e_k(P_k,T_k)=\rho_k\frac{P_k+\gamma_k \; P_{\infty,k}}{P_k+ P_{\infty,k}}c_{v_{k}}\; T_k +\rho_kq_k \\
g_{k}(P_k,T_k)=(\gamma_k \; c_{v_{k}}-q_{k}^{'})T_{k}-c_{v_{k}}T_{k}ln \left[\frac{T_k^{\gamma_k}}{(P_k+P_{\infty,k})^{\gamma_k-1} } \right]+q_k\\
\tau_k=\frac{(\gamma_k-1)c_{v_{k}}T_k}{P_k+P_{\infty,k}}
\end{cases}
\end{equation}
Let us note that the polytropic coefficient $\gamma_k$ here is the constant ratio of specific heat capacities of phase $k$, and $c_{v_{k}}$ and $P_{\infty,k}$ denote respectively the specific heat  at constant volume and a constant reference pressure for phase $k$. The specific energy of the fluid at a given reference state for phase $k$ is denoted by $q_k$, while $\tau_k=\frac{1}{\rho_k}$ corresponds to the specific volume.

The mixture temperature reads
\begin{equation}
T(P, \alpha_l,\rho)=\left(\frac{\alpha_l\rho_l{c_{v_l}}\; (\gamma_l -1)}{ P+P_{\infty,l}   }+\frac{(1-\alpha_l)\rho_g{c_{v_g}}
\;(\gamma_g -1) }{  P+P_{\infty,g} }\right)^{-1}
\end{equation}
while the mixture internal energy is
\begin{equation}
\begin{split}
e(T,P, \alpha_l)=&\alpha_l \rho_l \Big{(}\,{c_{v_l}}\; T \frac{P +\gamma_l\; P_{\infty,l}}{P+P_{\infty,l}}+\;q_l \Big{)}\\&
+(1-\alpha_l)\rho_g\Big{(}{c_{v_l}}\; T \frac{P+\gamma_g\; P_{\infty,g}}{P+P_{\infty,g}}+ \;q_g \Big{)}.
\end{split}
\end{equation}
As for the pressure, we consider \cite{le2014towards}
\begin{equation}
P(e, \alpha_l,\rho)=\frac{1}{2} (A_l+A_g-(P_{\infty,l}+P_{\infty_g}))+\sqrt{\frac{1}{4}(A_g-A_l-(P_{\infty,g}-P_{\infty,l}))^2+A_l\;A_g},
\end{equation}
where 
\begin{equation}
A_k=\frac{\rho_k\alpha_k(\gamma_k-1) c_{v_{k}}}{\rho_l\alpha_lc_{v_l}+\rho_g(1-\alpha_l){c_v}_g} \Big{(} e-(\alpha_l \; \rho_l \; q_l+(1-\alpha_l)\; \rho_g \;q_g) -P_{\infty,k}\Big{)}.
\end{equation}
We remark, that within this model, as usually done in the context of Baer-Nunziato type systems, the "pure" phase always contains a small amount of the other phase, s.t. $\alpha_k=1-\epsilon$, with $\epsilon\approx10^{-6}$, due to technicalities.\\

\section{Non-conservative Second-Order Explicit Residual Distributions Scheme}\label{PSIgalerkin}

The numerical approximation considered within this work follows the idea to solve the presented four-equation velocity-, pressure- and temperature-equilibrium system via a splitting method, which takes first into account only the hyperbolic problem given by \eqref{4_equation_convective} and only successively corrects the solution of the hyperbolic system by adding the source terms, such that the solution is updated with the mass transition of the phases as \eqref{4_equation_source}.

\begin{equation}
\begin{cases}
\partial_t (\alpha_1 \rho_1)+ \text{div}\;(\alpha_1 \rho_1 \; \textbf{u})=0\\[0.5em]
\partial_t (\alpha_2 \rho_2)+ \text{div}\;(\alpha_2 \rho_2 \;\textbf{u})=0\\[0.5em]
\partial_t ( \rho \;\textbf{u})+\text{div}\;( \rho\;\textbf{ u} \otimes \textbf{u} +P \,Id)=0\\[0.5em]
\partial_t e+ \textbf{u} \cdot \nabla e  + (e+P)\;\text{div}\; \textbf{u}=0\\
\end{cases} 
\label{4_equation_convective}
\end{equation}

\begin{equation}
\begin{cases}
\partial_t \alpha_1 \rho_1={\Gamma}  \\[0.5em]
\partial_t \alpha_2 \rho_2=-{\Gamma}  \\[0.5em]
\partial_t \rho \mathbf{u} = 0  \\[0.5em]
\partial_t e=0
\end{cases} 
\label{4_equation_source}
\end{equation}


In the 1D case, the set of equations \eqref{4_equation_convective} can be rewritten as 
\begin{equation}
\begin{cases}
\partial_t \U + \text{div}\, \mathbf{F}(\U) =0\\[0.1em]
\partial_t e+\textbf{u} \cdot \nabla e +(e+P) \text{div\,} \textbf{u}=0\\[0.1em]
\U(x,0) = \U^0(x),\\
\end{cases}
\label{system_4MF}
\end{equation}
on $\Omega \times [0,T]$, with $\U=[\alpha_1\rho_1,\; \alpha_2\rho_2 ,\; \rho\textbf{u}]^T$ and with the flux defined as
$$ \mathbf{F}(\U) = (\alpha_1\rho_1 u,\; \alpha_2\rho_2 u,\;\rho \textbf{ u} \otimes \textbf{u} +P \,Id)^T .
$$

\subsection{Approximation of the hyperbolic part}

\begin{figure}[H]
\centering
\includegraphics[scale=0.65]{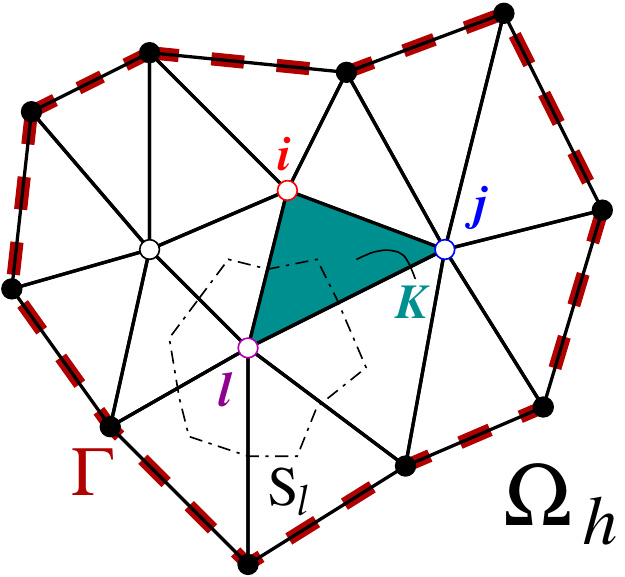}
\caption{Discretized domain $\Omega_h$ and its boundary $\Gamma$ for two dimensions ($d=2$) and simplexes. }
\label{Step1_RD}
\end{figure}

Inspired by the work of \cite{Ricchiuto2010}, \cite{BacigaluppiNICFD2016} and \cite{AbgrallFVCA8}, we have designed the discretiztion in time for \eqref{4_equation_convective} with a second-order explicit (Runge-Kutta) method. This time-stepping approach is used in combination with the Residual Distribution discretization in space, which has been accurately presented in some recent work \cite{AbgrallHO2018,BacigaluppiMood2018}.\\\\
Let us start by splitting the spatial domain $\Omega$ in $N_e$ conformal non-overlapping elements with characteristic length $\Delta x$ or $h$, and denote the set of all
the elements by ${\Omega}_h$,  as for example in Figure \ref{Step1_RD}. 
$K$ refers to a generic element of the tessellation $\Omega_h$, and  $\Gamma=\partial {\Omega}_h$ represents the boundary of the domain's problem.\\
Within each time step $[t_n, t_{n+1}]$ we consider $M$ subintervals, so that $t_n = t_{n,0} < t_{n,1} < \dots < t_{n,m} < \dots < t_{n,M} = t_{n+1}$.

Next, for each subinterval $[t_{n,m},t_{n,m+1}]$, we denote the approximated solution at the $m$-th substep $t_{n,m}$ as $\U_h^{n,m}$ and the solution at $t_n$ as $\U_h^n$.

The proposed time-stepping is recast as a predictor-corrector approximation with $M=2$. For any degree of freedom $\sigma \in K$ we have an

\begin{itemize}
\item initialization: $\U_{\sigma}^{(n,1)}=\U_{\sigma}^{{(n,0)}}=\U_{\sigma}^n$;

\item loop for $m=0,1$:  
\begin{equation}
|S_{\sigma}| \dfrac{ \U_{\sigma}^{(n,m+1)}-U_{\sigma}^{{(n,m)}}}{\Delta t_{n,m}}+\sum_{K | \sigma \in K} \phi_{\sigma}^K(\U^{(n,m)}_h,\U^{{(n,0)}}_h)=0,
\label{corrector_4MF}
\end{equation}
where $ \phi_{\sigma}^K$ is a so-called nodal residual which is detailed in the next paragraph and $|S_{\sigma}|$ is the area of the median dual cell $S_{\sigma}$ obtained by joining the gravity centres of the cells with the midpoints of the edges meeting in the node $\sigma$ (see cf. \cite{Ricchiuto2007} and Figure \ref{Step1_RD}).
\end{itemize}

The idea of the prediction step is thus to be first order accurate and to be given by a flux difference. The second-order in time is then achieved with the correction step, which takes the obtained prediction approximation as the previous sub-timestep solution.  The second-order in space is guaranteed by a correct choice of the nodal residual term $\phi_{\sigma}^K$ in \eqref{corrector_4MF}.\\

In the next paragraphs, we provide details on the actual choice of $\phi_{\sigma}^K$ (cf. Section \ref{Sec:MOOD}) and the conservative approach for the treatment of the non-conservative formulation of system \eqref{system_4MF} (cf. Section \ref{Sec:cons_noncons}).

  \subsection{Optimization via an ``a posteriori'' limiting}\label{Sec:MOOD}
 
 It is well known that in literature several benchmark problems related to multiphase flow systems are extremely difficult to be reproduced as they require an extreme robustness of the scheme. 
 This might be the case, for example, when very high gradients of pressures or velocities are defining the set-up of the problem.
On one hand, a highly accurate scheme with little or no dissipation mechanisms looses in terms of robustness across strong interacting discontinuities for  benchmark problems designed for extremely severe initial conditions.
On the other hand, adopting a dissipative, first-order formulation  would not allow to retrieve the desired accuracy in smooth regions.
To overcome  these outlined issues, we consider the proposed scheme of \cite{BacigaluppiMood2018}, which combines different numerical schemes to retrieve high accuracy in smooth regions of the flow, while, at the same time, ensuring robustness and a non-oscillatory behaviour in regions of steep gradients.  This idea is also known as the Multidimensional Optimal Order Detection (MOOD) approach (\hspace{1sp}\cite{Clain2011}, \cite{Diot2012}) and allows to introduce local dissipation if needed.
Nevertheless, this strategy differs from those well known available in literature, as shall explained in the following.
We consider $\phi_{\sigma}^K$ of equation \eqref{corrector_4MF} rewritten as
\begin{equation}
{\phi_\sigma^{K,s}(U)} =\int_K \varphi_{\sigma} \partial_t U_{\sigma} \,d\mathbf{x}+   \phi_{\sigma,\mathbf{x}}^{K,s}(U), 
\label{phi_compos}
\end{equation}
where we have dropped for simplicity the subscripts denoting the time dependency, s.t. $\phi_{\sigma}^{K,s}(U)=\phi_{\sigma}^{K,s}(\U^{(n,m)}_h,\U^{{(n,0)}}_h)$.\\

In the following, we consider a Finite Element-based approach, but it has been shown in \cite{Abgrall2001} that residual based configurations may be rewritten in terms of Finite Volumes or Discontinuous Galerkin schemes, with the advantage that Riemann solvers can be incorporated, but it is not mandatory. \\
Considering a finite element type setting, we thus consider our approximation space to be given by globally continous piecewise polynomials of degree $k$, s.t. $V_h=\{U\in C^0(\Omega_h), U|_{K} \in \mathcal{P}^k, \forall K \in \Omega_h\}$.
The $\varphi_{\sigma}$ in \eqref{phi_compos} denotes a basis function of choice, as, for instance, given by Lagrangian polynomials, and such that $
U_h= \sum_{K \in {\Omega_h}} \sum_{\sigma \in K} U_\sigma\varphi_\sigma{\Big|_{K}}.$\\

\noindent The optimization strategy is designed for the current work as follows:
\begin{itemize}
\item[1.] For each degree of freedom $\sigma \in K$ we start with the solution at time $t_{n}$ and 
\item[2.] compute the candidate solution $\U_h^{K,s,n+1}$ for time $t_{n+1}$.
\item[3.] The candidate solution is checked over a series of admissibility criteria, as listed in Section \ref{Detection}.

\begin{itemize}
\item[a)] If the candidate solution does not fulfill one of the criteria in a cell $K$, we use locally on the troubled cell a more dissipative scheme, repeating steps 2-3. We go to step 4, in case we have reached a ``parachute'' scheme, typically of first order of accuracy. 

\item[b)] If the candidate solution does fulfill all the set admissibility criteria, we go to step 4.

\end{itemize}
\item[4.] We accept our candidate solution to be  $\U_h^{K,n+1}=\U_h^{K,\text{s},n+1}$.

\end{itemize}

In this work we loop two times over the steps 3-4, i.e. we consider in total two different schemes, $s=0,1,2$, including the parachute scheme to compute our candidate solution. At every violation of the admissibility criteria we impose $s=s-1$ locally on the troubled cell.
The scheme considered in this work are  
\begin{itemize}
\item $s=2$: ``stabilized Galerkin scheme'' of \cite{BacigaluppiMood2018}
\begin{equation} 
\begin{split}
\phi_{\sigma,\mathbf{x}}^{K,2}(U)=&\int_{\partial K}\varphi_{\sigma} \mathbf{F}(U)\cdot \mathbf{n}\,d\Gamma -\int_K \nabla \varphi_{\sigma}\cdot \mathbf{F}(U)\,d \mathbf{x} + \phi_{\sigma,\mathbf{x}}^{K,Jump}(U)
 \end{split}
\label{galerk_stab}
\end{equation}
 
\item $s=1$: ``stabilized, blended Rusanov scheme'' of \cite{AbgrallHO2018,BacigaluppiMood2018}
\begin{equation}
 \phi_{\sigma,\mathbf{x}}^{K,1}(U)=\phi_{\sigma,\mathbf{x}}^{K,\star}(U) + \phi_{\sigma,\mathbf{x}}^{K,Jump}(U),
\label{RusPsi_stab}
\end{equation}
with $\phi_{\sigma,\mathbf{x}}^{K,\star}(U) $ defined in paragraph \ref{Psi_Rus_comment}.

\vspace{0.1cm}
\item $s=0$: ``Rusanov scheme''
\begin{equation}
\phi_{\sigma,\mathbf{x}}^{K,0}(U)=\int_{\partial K}\varphi_{\sigma} \mathbf{F}(U)\cdot \mathbf{n}\,d\Gamma -\int_K \nabla \varphi_{\sigma}\cdot \mathbf{F}(U)\,d \mathbf{x} 
     +\alpha(\widetilde{U_{\sigma}}-\overline{U}^K).
\label{Rus}
\end{equation}
See paragraph \ref{Rus_comment} for a comment on the terms.
\end{itemize}
The stabilization for $s=1,2$ is expressed as $ \phi_{\sigma,\mathbf{x}}^{K,Jump}(U)=
 \theta h_e^2 \int_e [\nabla U]\cdot [\nabla \varphi_{\sigma}]\,d\Gamma $.
\subsubsection{A comment on the  Rusanov scheme ($s=0$)}\label{Rus_comment}
The Rusanov's scheme, also known as the local Lax-Friedrich's scheme, is in our case the parachute scheme, i.e. a first order approach. In particular it considers in our predictor-corrector approach the $\widetilde{U_{\sigma}}=\frac{U_{\sigma}^{n,m}+U_{\sigma}^{n,0}}{2}$  and $\overline{U}^K=\frac{1}{N_{DoF}} \Big{(}\sum_{\sigma \in K} \frac{1}{2}\big{(}U_{\sigma}^{n,m}+\U^{n,0}_{\sigma}\big{)} \Big{)}$. Here $N_{DoF}$ represents the number of degrees of freedom within one cell $K$.\\ Further, 
$\alpha_K$  allows to add numerical diffusion to the approximation in order to guarantee a monotone solution and is in the form of a characteristic velocity written as $$\alpha_K = \max\limits_{\sigma\in K} \bigg ( \rho_S\Big ( \nabla_{\U}\mathbf{F}(\U_h)\cdot \nabla\varphi_\sigma \Big )\bigg ).$$
To determine the spectral radius $ \rho_S\Big ( \nabla_U\mathbf{F}(U_h)\cdot \nabla\varphi_\sigma \Big )$ within a cell, we have used the arithmetic averages over one cell, such that $\alpha_K=\,max(|\lambda_{\sigma}^K(\widetilde{\U}_{\sigma})|) $. Note that $\lambda^K_{\sigma}$  represent the eigenvalues of the system.\\

\subsubsection{A comment on the stabilized, blended Rusanov scheme ($s=1$)}\label{Psi_Rus_comment}
For the sake of allowing the reproducibility of the presented method, we detail hereafter the technical choices behind the stabilized, blended Rusanov scheme ($s=1$).
To do so, we will in the following denote $\phi_{\sigma}^{K,0}(U)$ given by \eqref{phi_compos} with \eqref{Rus} as $\phi_{\sigma}^{K,Rus}(U)=\phi_{\sigma}^{K,Rus}(\U^{(n,m)}_h,\U^{{n,0}}_h)$, to avoid any possible misunderstanding.
Scheme $s=1$ considers a blending of the Rusanov scheme that allows for second-order accuracy. This relies on the decomposition in local characteristic variables of the first-order residuals such that, instead of $\phi_{\sigma}^{K,Rus}(U)$, one considers $\phi_{\sigma}^{K,\star}(U)$.\\
 \noindent To compute $\phi_{\sigma}^{K,\star}(U)$, we proceed as follows.
\begin{itemize}
\item 
We first consider the definition of the residual according to equation \eqref{Rus}

\item Then we consider the eigen-decomposition of $\mathbf{A}_\bn=\mathbf{A}(\widehat{\U}_K)n_x+\mathbf{B}(\widehat{\U}_K)n_y$ (in 2D, while in 1D it would simply be $\mathbf{A}(\widehat{\U}_K)$. The matrices $\mathbf{A}$ and $\mathbf{B}$ are the Jacobians of the $x$- and $y$- component of the flux $\mathbf{F}$ with respect to the state $\widehat{\U}_K$. Here, the vector $\bn$ is a unit vector in the direction of the velocity field when it is nonzero, or any arbitrary direction otherwise. Of course, this direction is not relevant in one dimension. The right eigenvectors of $\mathbf{A}_\bn$ are denoted by
$\{\mathbf{r}_\xi\}$. We denote by $\{\mathbf{\ell}_\xi\}$ the left eigenvectors of $\mathbf{A}_\bn$, so that any state $X$ can be written as:
$X=\sum_\xi \mathbf{\ell}_\xi(X)\mathbf{r}_\xi.$

\item Then we decompose, for any degree of freedom $\sigma$:
$$\phi_{\sigma}^{K,Rus}(U) =\sum_{\xi} \mathbf{\ell}_{\xi}\big ( \phi_{\sigma}^{K,Rus}(U)\big ) \mathbf{r}_{\xi}$$
We note that for any $\xi$,
$$\mathbf{\ell}_\xi\big ( \phi_\mathbf{x}^{K}\big )=\sum_{\sigma \in K} \mathbf{\ell}_\xi\big ( \phi_{\sigma}^{K,Rus}(U)\big ),$$
where $\phi_\mathbf{x}^{K} = \sum_{\sigma \in K}\phi_{\sigma,}^{K,Rus}$ is the total residual.

\item The next step is to define $\phi_{\sigma}^{K,\star}(U)$ as:
\begin{subequations}\label{grosbeta}
\begin{equation}
\label{star}
\phi_{\sigma}^{K,\star}(U)=\sum_\xi  \mathbf{\ell}^{\star}_\xi\big ( \phi^{K}_{\sigma}\big )\mathbf{r}_\xi
\end{equation}
where:
\begin{equation}
\label{betastar}
\mathbf{\ell}^{\star}_\xi\big ( \phi^{K}_{\sigma}\big )=(1-\Theta_\xi)\overline{\phi}_{\sigma}^{K,Rus}(U)
+\Theta_\xi \mathbf{\ell}_\xi\big ( \phi_{\sigma}^{K,Rus}(U)\big )
\end{equation}
with
\begin{equation}
\label{theta}
\Theta_\xi=\dfrac{\big | \mathbf{\ell}_\xi\big (\phi_{\sigma}^{K,Rus}(U)\big )\big |}{\sum\limits_{\sigma'\in K}\big | \mathbf{\ell}_\xi\big ( \phi_{\sigma'}^{K,Rus}(U)\big )\big |}
\end{equation}
and
\begin{equation}
\overline{\phi}_{\sigma}^{K,Rus}(U)=\dfrac{ 
\max \Big( \frac{\mathbf{\ell}_\xi\big ( \phi_{\sigma}^{K,Rus}(U)\big )}{\mathbf{\ell}_\xi\big [ \phi_{\mathbf{x}}^{K}\big ]},0\Big)
}
{\sum\limits_{\sigma'\in K}\max \Big( \frac{\mathbf{\ell}_\xi\big ( \phi_{\sigma'}^{K,Rus}(U)\big )}{\mathbf{\ell}_\xi\big [ \phi_{\mathbf{x}}^{K}\big ]},0\Big)}
\end{equation}
\end{subequations}
Note that $\Theta_\xi\in [0,1]$.
This guarantees that the scheme is second-order in time and space and (formally) non-oscillatory, see \cite{Ricchiuto2010,Abgrall2006,DeSantis2014} for more details. 
\end{itemize}

\subsubsection{Detection procedure for the ``a posteriori'' limiting} \label{Detection}
The detection criteria responsible to mark the approximation as acceptable are based on physical/modelling and numerical considerations.
The cell will be thus flagged as 'good' if its candidate solution fulfills all detection criteria
and it is not a direct neighbour  of a bad cell. 
These criteria reads 
\begin{equation}
\label{eq:bad_4MF}
\mathcal{B} = \left\{  K\in \Omega_h, \; \text{s.t.} \;
  \left( \text{PAD}_K \times \text{CAD}_K \times \text{P}_K  \times\text{NAD}_K = 1 \right) \; \text{or} \; 
  \left( \exists K' \in \mathcal{V}(K), \; K' \in \mathcal{B} \right)  \right\}.
\end{equation}

More precisely, in case of our four-equation model, the considered variables we first check for positivity are $\rho$ and the pressure $P$ at each degree of freedom in the 
cell, that is, if the cell $K$ fulfils the 
\begin{itemize}
\item Physical Admissibility Detection criteria
\begin{equation}
 \label{eq:PAD_4MF}
\text{PAD}_K= \left\{ 
    \begin{array}{lll}
      1 & \text{if} & \forall \sigma \in K, \; \; \rho_{\sigma}^{\star,n+1}<0 \;\text{or} \; P_{\sigma}^{\star,n+1}<0 \\
      0 & \text{else}. & 
    \end{array}\right.
\end{equation}
\end{itemize}

To prevent the numerical approximation to be Not-A-Number (NaN) or Infinity (Inf), we perform the check via the
\begin{itemize}
\item Computational Admissibility Detection criteria for all variables
\begin{equation}
 \label{eq:CAD_4MF}
 \text{CAD}_K = \left\{ 
    \begin{array}{lll}
      1 & \text{if} & \exists \sigma \in K, \; \; {\U}_\sigma^{\star,n+1}=\text{NaN}  \; \text{ or } \; {\U}_\sigma^{\star,n+1}=\text{Inf},\\
      0 & \text{else}. & 
    \end{array}\right.
\end{equation}
\end{itemize}
In case we are within a plateau area, we make sure to not break that area by applying a 
\begin{itemize}
\item Plateau Detection criteria
\begin{equation}
\label{eq:Plateau_4MF}
\text{P}_K = \left\{ 
    \begin{array}{lll}
      0 & \text{if} & \exists \sigma \in K,\quad |M^n-m^n| \geq \mu^3 ,\\
      1 & \text{else}. & 
    \end{array}\right.
\end{equation} 
    \end{itemize}
    The bounds are defined by
\begin{equation}
M^n = \max_{K' \in \mathcal{V}(K),  \sigma \in K' } \left( \U_\sigma^{n} \right),  \qquad
m^n = \min_{K' \in \mathcal{V}(K),  \sigma \in K' } \left( \U_\sigma^{n} \right), 
\end{equation}
where the neighbourhood $\mathcal{V}(K)$ is the set of cells surrounding $K$, and the relaxed parameter is
given by $\mu=|K|^{1/d}$, with $d$ the size of the considered dimensions within our problem set.

Finally, to prevent the approximation from numerical oscillations we adopt the
\begin{itemize}
\item Numerical Admissibility Detection criteria
\begin{equation} \label{eq:NAD_4MF}
\text{NAD}_K = \left\{ 
    \begin{array}{lll}
      1 & \text{if} & \text{  DMP}_K=1 \text{ and SE}_K=1\\
      0 & \text{else.} & 
    \end{array}\right.
\end{equation}
\end{itemize}
This criteria has two major steps: in case  the first one is activated,  it is necessary that also the second one is activated, in order to state that a cell is troubled at all. The first one is a so-called

\begin{itemize}

\item[1.] Relaxed Discrete Maximum Principle (DMP) criteria 
\begin{equation}
\text{DMP}_K = \left\{ 
    \begin{array}{lll}
      0 & \text{if} &  m^n<\U_\sigma^{\star,n+1} < M^n\\
      1 & \text{else}, & 
    \end{array}\right.
\end{equation}
\end{itemize}
In case the cell is marked with DMP$_K=1$, we check for the second criterion, i.e.  the
\begin{itemize}
\item[2.] Smoothness Extrema Criteria (SE),  inspired by \cite{Vilar2018,Wang2009,Kuzmin2010}. This second check allows to exclude the possibility of a mistakenly flagged cell, as, could happen, for example, in case of natural oscillations with a coarse mesh (c.f.  the Shu-Osher benchmark in \cite{BacigaluppiMood2018}).
\end{itemize}
This technique is based on a linearised version of the numerical spatial derivative.  In 1D, this check considers the value of $\hat{\alpha}_{\sigma}=\min(\hat{\alpha}_L, \hat{\alpha}_R)$. In case,   $\hat{\alpha}_\sigma<1$ for $\sigma \in K$ on $K$, the candidate solution is recomputed with a more dissipative scheme.
As for the definition of the $\hat{\alpha}_L$, we recast
\begin{equation}
\hat{\alpha}_L=\left\{ 
    \begin{array}{lll}
    \min\left(1, \dfrac{\U_{\max,L}-\overline{\partial_{\mathbf{x}} \U}_\sigma^{\star,n+1}}{\overline{\delta\U}_L^{\star,n+1}-\overline{\partial_{\mathbf{x}} \U}_\sigma^{\star,n+1} }\right), & \text{if} & \overline{\delta\U}_L^{\star,n+1}> \overline{\partial_{\mathbf{x}} \U}_\sigma^{\star,n+1}\\
    & & \\
     \min\left(1, \dfrac{\U_{\min,L}-\overline{\partial_{\mathbf{x}} \U}_\sigma^{\star,n+1} }{\overline{\delta\U}_L^{\star,n+1}-\overline{\partial_{\mathbf{x}} \U}_\sigma^{\star,n+1} }\right), & \text{if} & \overline{\delta \U}_L^{\star,n+1}< \overline{\partial_{\mathbf{x}} \U}_\sigma^{\star,n+1}\\
      & & \\
    1 & else &
      \end{array}\right.  \label{alpha_mood}
\end{equation}
and analogously $\alpha_R$.\\
The definitions introduced in \eqref{alpha_mood} are $\U_{\min \backslash \max,L}= \min \backslash \max \left( \overline{\partial_{\mathbf{x}} \U}_{\sigma-1}^{\star,n+1},
 \overline{\partial_{\mathbf{x}} \U}_{\sigma}^{\star,n+1}\right)$ and $\U_{\min \backslash \max,R}= \min \backslash \max\left( \overline{\partial_{\mathbf{x}} \U}_{\sigma}^{\star,n+1},
 \overline{\partial_{\mathbf{x}} \U}_{\sigma+1}^{\star,n+1}\right)$. 
 In one-dimension, further, we have
$$ \overline{\delta \U}_L^{\star,n+1}=
 \overline{\partial_{x} \U}_\sigma^{\star,n+1} -  \frac{\Delta x}{2}\overline{\partial_{xx} \U}_\sigma^{\star,n+1}, \quad \overline{\delta \U}_R^{\star,n+1}=
\overline{\partial_{x} \U}_\sigma^{\star,n+1} +  \frac{\Delta x}{2}\overline{\partial_{xx} \U}_\sigma^{\star,n+1}.$$
For the 2D definitions, the reader might refer to \cite{BacigaluppiMood2018}.

We remark, that for the plateau and numerical admissibility criteria, we verify the conditions on both the candidate solutions $\rho$ and $Y_1$ on $\sigma$  at time ${\star,n+1}$.

\subsection{A conservative approach for the non-conservative formulation}\label{Sec:cons_noncons}

In order to solve \eqref{4_equation_convective}, the scheme applied to the non-conservative equation has to be modified appropriately. 
In the present work, following \cite{AbgrallCAF2017}, to treat the equation written in terms of the internal energy
\begin{equation}
\partial_t e+ \textbf{u} \cdot \nabla e  + (e+P)\;\text{div}\; \textbf{u}=0, \label{noncons_eq}
\end{equation}
we first solve \eqref{noncons_eq} according to the chosen scheme to build a $\phi_{\sigma}^{K,s}(U)$, with $s=0,1,2$, according to either \eqref{RusPsi_stab} or \eqref{Rus}.
This means, considering e.g. $s=0$, that we will recast
\begin{equation}\label{Rusanov_noncons_4MF}
\begin{split}
\phi_{\sigma,e}^{K,Rus}=\frac{1}{N_{DoF}} \int_K \bigg(  
\frac{e_{h}^{(n,m)}-e_h^{n,0}}{\Delta t_{n,m}}&+
u_h\cdot \nabla e_h+(e_h+P_h )\text{ div }u_h\bigg ) d\mathbf{x}\\& + \alpha_K\big (\dfrac{ e_\sigma^{(n,m)}-e_\sigma^{n,0}}{2}-\overline{e}_K\big ).
\end{split}
\end{equation}

As, \eqref{Rusanov_noncons_4MF} does not preserve the conservation of the total energy. To allow the method to get relevant weak solutions, the residual of the internal energy $\phi_{\sigma}^{K,s}(U)$   is corrected via  
\begin{equation}
\widetilde{\phi_{\sigma,e}^{K,s}} =\phi_{\sigma,e}^{K,s}+r_{\sigma}^K
\label{phiE_tilde_4MF}
\end{equation}
i.e., in order to get local conservation of the energy and following \cite{AbgrallCAF2017}, we impose
\begin{equation}
\begin{split}
\sum\limits_{\sigma\in K}\phi_{\sigma,e}^{K,s}+\sum\limits_{ \sigma \in K}r_{\sigma}^K=\phi^{E,K}&- \sum\limits_{\sigma \in K}\frac{\mathbf{u}_{\sigma}^{n,m+1}+\mathbf{u}_{\sigma}^{n,m}}{2}\cdot 
 \phi_{\sigma \mathbf{m}}^{K,s}\\&+\frac{1}{2}\sum\limits_{\sigma\in K}\mathbf{u}_{\sigma}^{n,m}\cdot \mathbf{u}_{\sigma}^{n,m+1}\cdot\phi_{\sigma,\rho}^{K,s}.
 \end{split}
 \label{phiE_tilde_extensive_4MF}
\end{equation} 
To summarize, the spatial approximation for both the density $\phi_{\sigma,\rho}^{K,s}$ and momentum equations  $\phi_{\sigma,m}^{K,s}$ is chosen according to \eqref{Rus},  while the energy is approximated via the $\widetilde{\phi_{\sigma,e}^{K,s}}$ nodal residual according to \eqref{phiE_tilde_extensive_4MF}.\\

\begin{figure}
\centering
\subfigure[Finite Volume Sketch]{\includegraphics[scale=0.3]{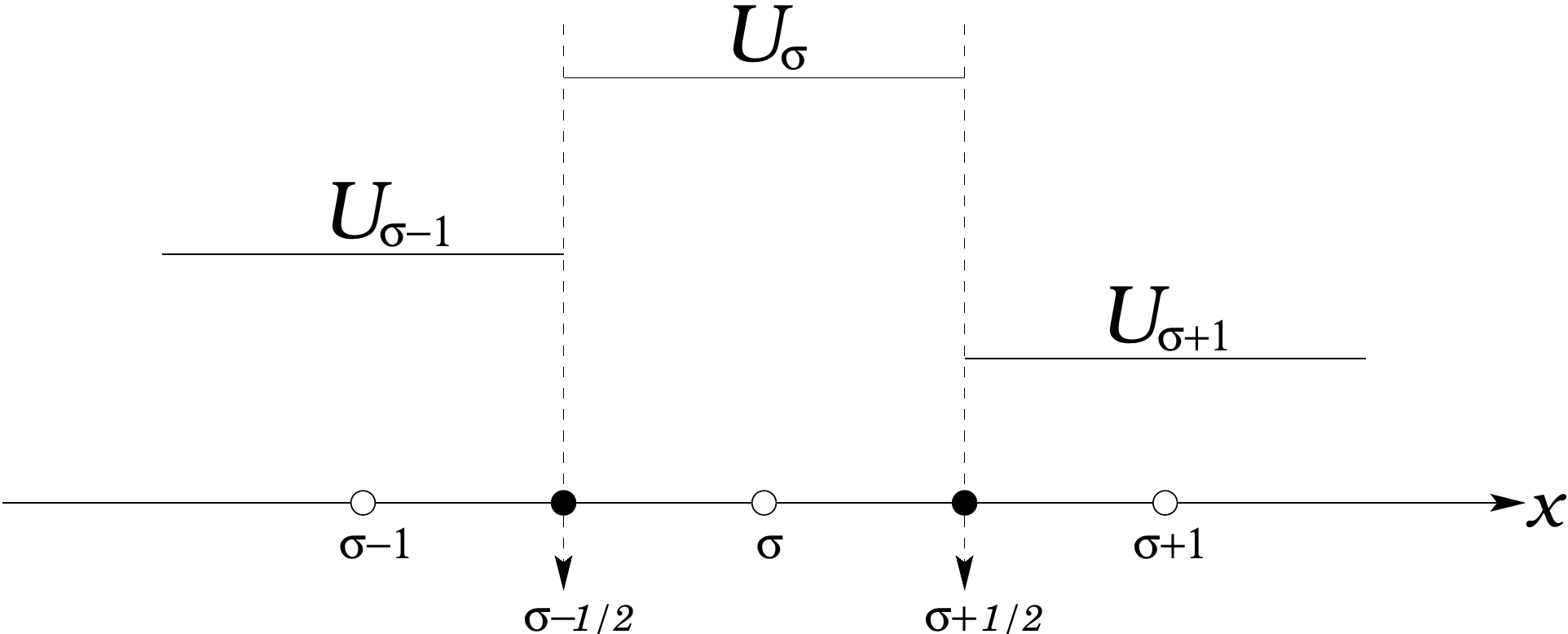}\label{FV_sketch}}
\vspace{2.5pt}
\subfigure[RD Sketch]{\includegraphics[scale=0.3]{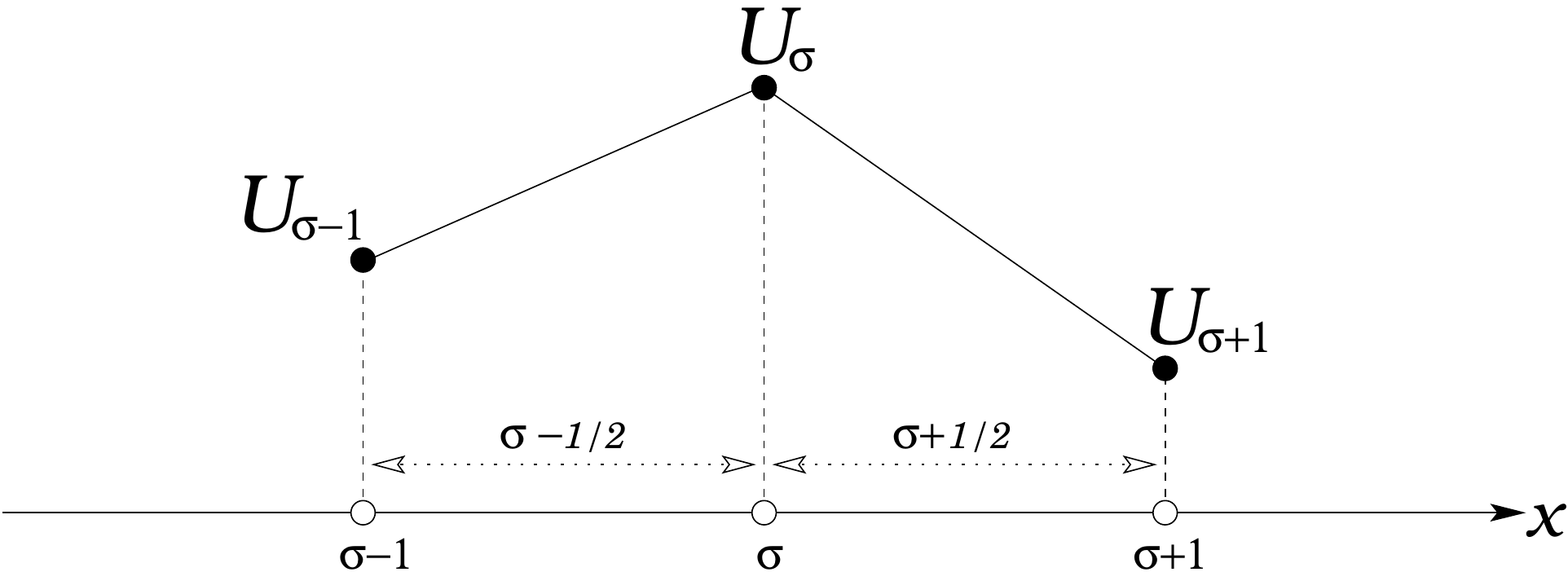}\label{RD_sketch}}
\caption{Sketches used for the definitions of Equation \eqref{FV_classicscheme} (left) and for Equation \eqref{RD_FVform} (right).}
\label{RD_FS_sketch}
\end{figure}

The idea behind \eqref{phiE_tilde_extensive_4MF} lies in the concept of the RD formulation, which can be recast in terms of Finite Volumes. Indeed, considering a classical finite volume scheme 
\begin{equation}
\U_{\sigma}^{n+1}=\U_{\sigma}^n-\frac{\Delta t}{\Delta x}\big (\hat{\mathbf{F}}_{{\sigma}+\frac{1}{2}}-\hat{\mathbf{F}}_{{\sigma}-\frac{1}{2}}), \label{FV_classicscheme}
\end{equation}
defined on a simple conservation law $\quad \partial_t \U+\nabla \cdot  \mathbf{F}(\U)=0$, and with reference to Figure \ref{FV_sketch} we can rewrite \eqref{FV_classicscheme} by adding and subtracting the term $ \mathbf{F}(\U_{{\sigma}})$, such that
\begin{equation}
\U_{\sigma}^{n+1}=\U_{\sigma}^n-\frac{\Delta t}{\Delta x}\left[ \big (\hat{\mathbf{F}}_{{\sigma}+\frac{1}{2}}-\mathbf{F}(\U_{\sigma})\big) + \big(\mathbf{F}(\U_{{\sigma}})-\hat{\mathbf{F}}_{{\sigma}-\frac{1}{2}}\big)\right].
\label{FV_classicreform}
\end{equation}

\noindent Denoting the definition of the flux differences in terms of fluctuation spitting, i.e. residuals,  we define 
 $\phi_\sigma^{\sigma +\frac{1}{2}}=\hat{\mathbf{F}}_{\sigma+\frac{1}{2}}-\mathbf{F}(\U_{\sigma})$,
 $\phi_\sigma^{\sigma -\frac{1}{2}}=\mathbf{F}(\U_{\sigma})-\hat{\mathbf{F}}_{\sigma-\frac{1}{2}}$ and
 $\phi^{\sigma +\frac{1}{2}}=\int_{x_\sigma}^{x_{\sigma+1}} \nabla \cdot  \mathbf{F} dx.$

\noindent This allows to reformulate \eqref{FV_classicreform}, with reference to Figure \ref{RD_sketch}, as 
\begin{equation}
\U_{\sigma}^{n+1}=\U_\sigma^{n}-\frac{\Delta t}{\Delta x}(\phi_\sigma^{\sigma+\frac{1}{2}}+\phi_\sigma^{\sigma-\frac{1}{2}})
\label{RD_FVform}
\end{equation}

\noindent  For \eqref{RD_FVform}, the conservation is recovered, if $\forall$ elements $[x_\sigma, x_{\sigma+1}]$:
\begin{equation}
\phi_\sigma^{\sigma+\frac{1}{2}}+\phi_{\sigma+1}^{\sigma+\frac{1}{2}}=\mathbf{F}(\U_{\sigma+1})-\mathbf{F}(\U_\sigma).
\label{RD_FVconservation}
\end{equation}

Consider now  the definition of $\Delta g=g^{n,m+1}-g^{n,m}$ and equivalently $\Delta (gh)= g^{n,m+1}\Delta h+h^{n,m}\Delta g =\overline{g}\Delta h+\underline{h}\Delta g$, with
$\overline{gh}=\overline{g}\; \overline{h}$ and $\underline{gh}=\underline{g}\; \underline{h}$.

\noindent With these definitions, we reformulate the total energy in terms of
\begin{equation}
\begin{split}
\Delta E&=\Delta e+\frac{1}{2}\Delta (\rho u^2)=\Delta e+\frac{1}{2}\bigg( \overline{\rho u}\cdot \Delta u+\underline{u}\cdot \Delta  \mathbf{m}\bigg)\\&=\Delta e+\frac{\overline{u}+\underline{u} }{2}\cdot \Delta \mathbf{m}-\frac{1}{2}\overline{u}\cdot \underline{u}\; \Delta \rho.
\end{split} 
\label{Etot_rewritten}
\end{equation}

From the finite volume scheme,  considering $i \in N_e$ the cell index set, and defining the set $ \mathcal{V}_i$ of all indexes $j \in N_e$ such that elements $K_j$ share a common face such that $K_i \cap K_j\neq0$.
\begin{equation}
|C_i| \big ( V_i^{n+1}-V_i^n)+\sum_{j\in \mathcal{V}_i} \hat{\mathbf{f}}_{ij}=0,  \; \text{with}\;\delta \hat{f}:=\sum_{j\in \mathcal{V}_i} \hat{f}_{ij}, 
\end{equation}
we have the system approximated as 
\begin{equation}
\begin{cases}
&|C_i|(\rho_i^{n+1}-\rho_i^n)+\delta\hat{f}_\rho =0,\\&
|C_i|(\mathbf{m}_i^{n+1}-\mathbf{m}_i^n)+\delta\hat{f}_{\mathbf{m}} =0,\\
&|C_i|(e_i^{n+1}-e_i^n)+ \delta \hat{f}_e=0
\end{cases}\label{FV_systemreform}
\end{equation}

\noindent From \eqref{Etot_rewritten} and \eqref{FV_systemreform} we define the flux difference for the internal energy as
\begin{equation}
\delta \hat{f}_e=\delta \hat{f}_E-\frac{\overline{u}-\underline{u} }{2}\cdot \delta \hat{f}_{\mathbf{m}}+\frac{1}{2} \overline{u}\cdot \underline{u}\; \delta \hat{f}_\rho.\label{dfe_FV}
\end{equation}

The analogy between \eqref{FV_classicreform} and \eqref{RD_FVform} allows to define a residual for the total energy by simply reinterpreting \eqref{dfe_FV} as 

\begin{equation}
\Phi_{i,e}^{K}=\Phi_{i,E}^{K}+\frac{\overline{u_i}-
\underline{u_i} }{2}\cdot \Phi_{i,\mathbf{m}}^{K}+\frac{1}{2}\overline{u}_i\cdot \underline{u}_i\cdot\Phi_{i,\rho}^{K},\label{dphie_RD}
\end{equation} 
\noindent with $K$ referring to a generic element of the tessellation $\Omega_h$.

\noindent By construction we have 
\begin{equation}
\sum_{i \in K}\Phi_{i,e}^{K}=\Phi_e^K
\end{equation}
and the  sequence of solutions will converge to a weak solution.

To guarantee the conservation of the quantities in \eqref{dphie_RD},  the terms $\overline{u}$ and $\underline{u}$ are choosen specifically according to the sub-time step. In particular, for the prediction, they are defined as $ \overline{u}=u^{(n,1)}$, $\underline{u}=u^{(n,0)}=u^n$, while for the correction as $ \overline{u}=u^{(n,2)}$, $\underline{u}=u^{(n,1)}$.

\subsection{Treatment of the heat transfer source term}
Once an approximation to \eqref{4_equation_convective} has been obtained, the solution is corrected by adding the heat transfer sources terms \eqref{4_equation_source}. 
Let us start by doing some considerations on the source term $\Gamma=\theta(g_l-g_g)$ that
models the mass transfer. 
We have vaporization if the Gibbs free energy $g_l$ of the liquid phase
is greater than the Gibbs free energy  $g_v$ of the vapor phase, and 
condensation otherwise. In our numerical model this mass transfer process 
might be activated or not depending on the application problem. Moreover,
it might be activated at selected locations, for instance when the volume fraction is greater than some tolerance. When mass transfer is activated we assume that this 
process is instantaneous, hence  the relaxation parameter $\theta$ is considered to be infinite
(zero relaxation time).
Otherwise, if no mass transfer is activated, then $\theta=0=\Gamma$.

Mass transfer here is treated numerically by using the technique of 
 \cite{Pelanti2014}. Since we assume an instantaneous process, the state after
 Gibbs free energy relaxation can be found by imposing thermodynamic equilibrium conditions,
 this resulting in a system of algebraic equations to be solved for the unknown relaxed state. We denote hereafter by superscript $0$ the computed quantities obtained from the approximation of the hyperbolic system \eqref{4_equation_convective} and by $\star$ those 
corresponding to the thermodynamic equilibrium state determined by mass transfer \eqref{4_equation_source}.
First, we observe that during  mass transfer the mixture density, momentum and energy are invariant:
\begin{equation}
\begin{cases}
\rho^0=\rho^\star,\\
(\rho \u)^0=(\rho \u)^\star,\\
e^0=e^\star.
\end{cases}
\end{equation}
 Through the relations $\rho^0=\alpha_l^*\rho_l^*+\alpha_g^*\rho_g^*$, $e^0=\alpha_l^*e_l^*+\alpha_g^*e_g^*$, $\rho_{l,g}=\rho_{l,g}(P,T)$, $e_{l,g}=e_{l,g}(P,T)$,
it is possible to obtain a formulation that links the pressure to the temperature.

Indeed, considering \eqref{thermo_4mf} along the thermodynamic closure \eqref{closure_mf4}, one obtains a quadratic expression that links the pressure to the temperature:
 \begin{equation} 
 \begin{aligned}
&a_p(P^*)(T^*)^2+b_p(P^*)T^*+d_p(P^*)=0,& \\
&a_p(P^*)=\rho^0 {c_v}_l {c_v}_g( (\gamma_g-1)(P^*+\gamma_l P_{\infty, l})-(\gamma_l-1)(P^*+\gamma_g P_{\infty,g}), & \\
& b_p(P^*)=e^0( (\gamma_l-1){c_v}_l(P^*+P_{\infty,g})-(\gamma_g-1){c_v}_g(P^*+P_{\infty,l}))& \\
& \qquad \qquad +\rho^0( (\gamma_g-1){c_v}_g q_{l}(P^*+P_{\infty,l}-(\gamma_l-1){c_v}_lq_{g}(P^*+P_{\infty g}) ) ,&\\
& \qquad \qquad  +{c_v}_g (P^*+P_{\infty,g})(P^*+\gamma_g P_{\infty,g})-{c_v}_l(P^*+P_{\infty,l})(P^*+\gamma_l P_{\infty,l}), & \\
&d_p(P^*)=(q_{g}-q_{g})(P^*+P_{\infty,l})(P^*+P_{\infty,g}).  &
 \end{aligned}
 \label{P_T_mass} 
 \end{equation} 
Imposing the condition of chemical equilibrium  $g_l^*=g_g^*$ for the two phases and
by adopting a stiffened gas equation of state one obtains the pressure-temperature saturation curve that reads
\begin{equation} 
\begin{split}
\frac{{c_p}_l-{c_p}_g+q_g^{'}-q_l^{'}}{{c_p}_g-C_{v_l}}&+\frac{1}{T}\frac{q_{l}-q_{g}}{{c_p}_g-{c_v}_g}+\frac{{c_p}_g-{c_p}_l}{{c_p}_g-{c_v}_g}logT\\&+\frac{{c_p}_l-{c_v}_l}{{c_p}_g-{c_v}_g}log(P+P_{\infty,l})-log(P+P_{\infty,g})=0.
\end{split}
 \label{chemical_egality} 
 \end{equation} 
The equation \eqref{chemical_egality} is non-linear and can be solved for the equilibrium pressure, for example, by using the Newton-Rhapson Method.\\
After finding the pressure $P^*$ from \eqref{chemical_egality}, it is possible to obtain from  \eqref{P_T_mass} the corresponding value of the equilibrium 
temperature $T^*$, selecting the physically admissible solution of the two resulting roots. Then all the variables of the equilibrium state reached after mass transfer can be obtained. See \cite{Pelanti2014} for more details. 
In case a physically admissible solution is not found, it means that we have a complete evaporation or condensation and we impose a negligible quantity of mass fraction $Y_k=\frac{\alpha_k \rho_k}{\rho}$ for the vanishing phase $k$, i.e. in our specific case $\alpha_k=10^{-8}$ (c.f. \cite{Pelanti2014}).

\section{Conservative wave propagation scheme}\label{sec:cons}
 In order to cross-validate the four-equation model, we solve the  homogeneous portion of the conservative system \eqref{4eqsmodel_mass_c} through a wave propagation scheme \cite{rjl:fvbook} implemented within the \textsc{Clawpack} software \cite{LeVeque2011}. This approach is based on an alternative formulation of the Godunov-type schemes, namely the wave propagation formulation
(here in one dimension for simplicity)
\begin{equation}\label{CLaw}
U_\sigma^{n+1}=U_\sigma^{n}-\frac{\Delta t}{\Delta x} (\mathbf{A}^+ \Delta U_{\sigma-\frac{1}{2}}+\mathbf{A}^- \Delta U_{\sigma+\frac{1}{2}})
+\frac{\Delta t}{\Delta x}(\mathcal{F}_{\sigma-\frac{1}{2}}-\mathcal{F}_{\sigma+\frac{1}{2}}).
\end{equation} 
A Riemann solver is used to provide the Riemann solution structure
at cell interfaces $\sigma+\frac{1}{2}$ in terms of waves $W^l$
and corresponding propagation velocities $s^l$,   $l = 1,\ldots, M$,
with  $M$ denoting the number of waves of the chosen Riemann solver
($M=3$ for the HLLC solver). 
Then the left-going and right-going fluctuations in (\ref{CLaw}) 
 arising  from Riemann problems at cell interfaces $x_{\sigma+1/2}$  
 are defined as:
\begin{equation}\label{fluctu_Claw}
\mathit{A}^{\pm}\Delta U_{\sigma+\frac{1}{2}}= \sum_{l=1}^M (s_{\sigma+\frac{1}{2}}^l)^{\pm} \mathit{W}_{\sigma+\frac{1}{2}}^l.
\end{equation}
 Let us recall that the Riemann solution structure
$\bigg\lbrace\mathit{W}_{\sigma +\frac{1}{2}}^l,s_{\sigma +\frac{1}{2}}^l\bigg\rbrace_{{l=1,2,\ldots, M}}$
provided by the Riemann solver must satisfy 
$\Delta U =U_{\sigma+1}-U_{\sigma}=\sum_{l} W_{\sigma+\frac{1}{2}}^l$.
 Furthermore, for conservation laws we require  
$\Delta f(U) =f(U_{\sigma+1})-f(U_{\sigma})=\sum_{l}s^l_{\sigma +\frac{1}{2}} W^l_{\sigma +\frac{1}{2}}$.
 
\noindent The terms $\mathcal{F}_{\sigma \mp \frac{1}{2}}$ in (\ref{CLaw}) correspond to  correction 
fluxes to achieve (formal) second-order accuracy
\begin{equation}
\mathcal{F}_{\sigma+\frac{1}{2}}=\frac{1}{2}\sum_{l=1}^M |s_{\sigma+\frac{1}{2}}^l| (1- \frac{\Delta t}{\Delta x} |s_{\sigma+\frac{1}{2}}|)\bar{\mathit{W}}_{\sigma+\frac{1}{2}}^l.
\end{equation}
Here $\bar{\mathit{W}}_{\sigma+\frac{1}{2}}^l$ are a limited version of the
waves $\mathit{W}_{\sigma+\frac{1}{2}}^l$, obtained as
$$
\bar{\mathit{W}}_{\sigma+\frac{1}{2}}^l = \phi(\theta_{\sigma+\frac{1}{2}}^l) 
\mathit{W}_{\sigma+\frac{1}{2}}^l,
$$
where $\phi(\theta)$ is a limiter function. The smoothness indicator
$\theta$ can be defined as:
$$
\theta_{\sigma+\frac{1}{2}}^l =\frac{\mathit{W}_{I+\frac{1}{2}}^l
\cdot \mathit{W}_{\sigma+\frac{1}{2}}^l}{
\mathit{W}_{\sigma+\frac{1}{2}}^l\cdot \mathit{W}_{\sigma+\frac{1}{2}}^l},
$$
where $I=\sigma-1$ if $s^l_{\sigma+\frac{1}{2}}>0$ and 
$I=\sigma+1$ if $s^l_{\sigma+\frac{1}{2}}<0$.
Here a Minmod limiter is used, $\phi(\theta) = \max(0,\min(1,\theta))$

\begin{figure}[h]
\centering
\includegraphics[scale=0.27]{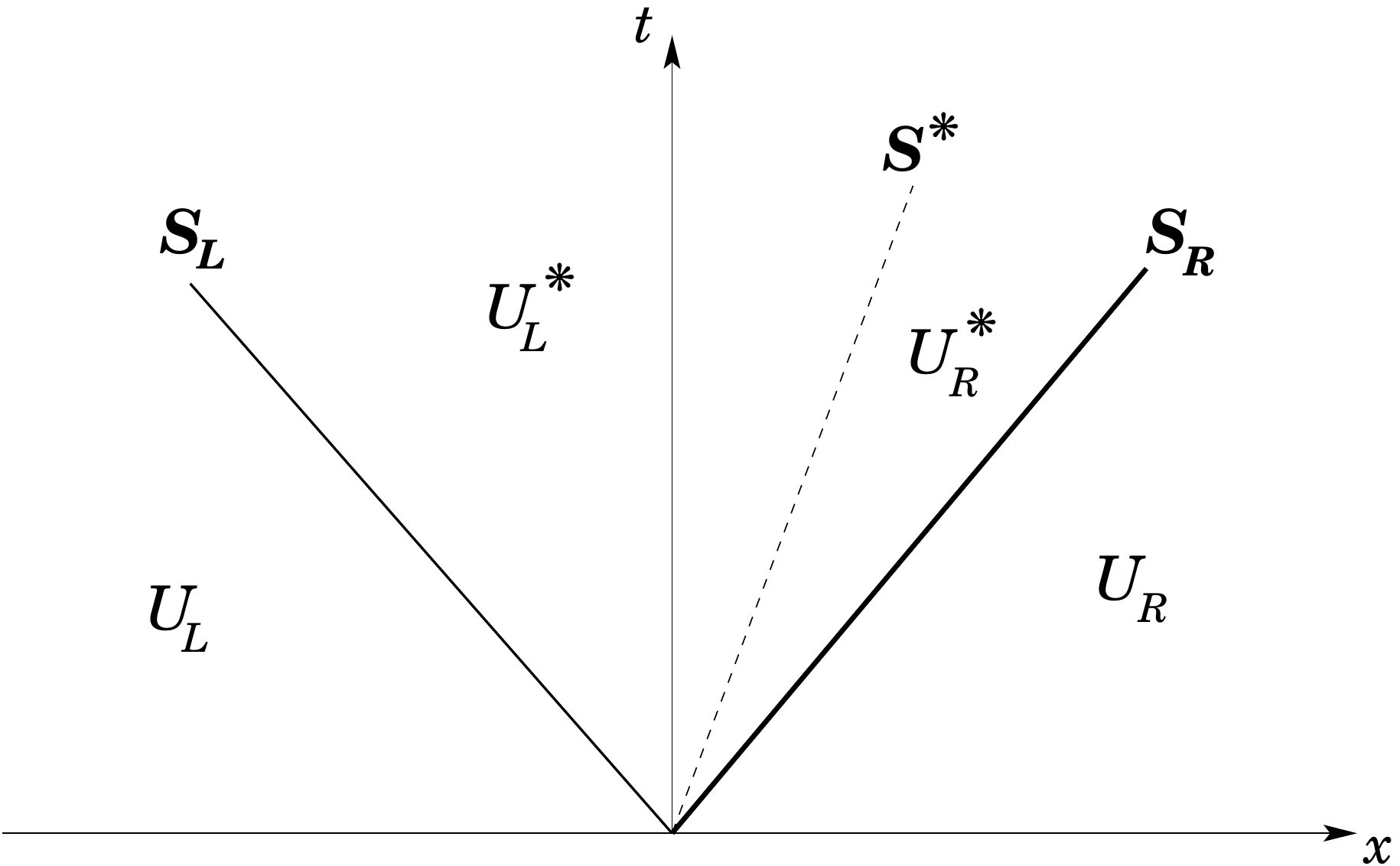}\caption{Representation of the HLLC wave structure solution for a Riemann problem. Each wave propagating respectively with speed $S_L$, $S^\star$ and $S_R$ represents a discontinuity.}
\label{HLLC_wave}
\end{figure}

\noindent In order to define the fluctuations $\mathit{A}^{\pm}\Delta U_{\sigma+\frac{1}{2}}$ for our four-equation system, i.e.\ to retrieve the Riemann solution structure $\bigg\lbrace\mathit{W}_{\sigma +\frac{1}{2}}^l,s_{\sigma +\frac{1}{2}}^l\bigg\rbrace_{l=1,2,\ldots, M}$, we chose a 
HLLC-type solver, which was first proposed in \cite{Toro1994}. Let us consider a Riemann problem with left and right data $U_L$ and $U_R$, respectively. 
The HLLC  solver consists of $M=3$  waves $\mathit{W}^l$ moving at  
speeds $s^1=S_L$, $s^2=S^\star$ and  $s^3=S_R$, and separating four constant states $U_L$, $U_L^\star$, $U_R^\star$ and $U_R$, as shown in Figure \ref{HLLC_wave}.

\noindent Following \cite{Davis1988}, we define the speed $S_L=\min \left((u-c)_R,(u-c)_L \right)$, 
$S_R=max((u+c)_R,(u+c)_L)$ 
and $$ S^\star=u^\star=\frac{P_R-P_L+\rho_{L} u_L(S_L-u_L)-\rho_{R} u_R(S_R-u_R)}{\rho_{L}(S_L-u_L)-\rho_{R}(S_R-u_R) }. $$

\noindent Furthermore, in order to obtain the middle states $U_L^\star$ and $U_R^\star$ for the \eqref{4eqsmodel_mass_c}, we follow \cite{SAUREL201653}. Setting $\xi=L,R$, we write
  \begin{equation} 
U_{\xi}^\star=
\begin{cases}
&\rho_{{\xi}}^\star=\rho_{{\xi}} \frac{u_{\xi}-S_{\xi}}{S^\star-S_{\xi}}, \\
&u_{\xi}^\star=u^\star=S^\star,\\
&P_{\xi}^\star=P_{\xi}+\rho_{{\xi}}\left( u_{\xi}-S_{\xi} \right) \left( u_{\xi}-S^\star \right),\\
&E_{{\xi}}^\star=E_{{\xi}}+\frac{P_{\xi} \left(u_{\xi}-S^\star \right)}{\rho_{\xi}\left(u_{\xi}-S_{\xi}\right)}-S^\star \left(u_{\xi}-S^\star \right),\\
&Y_{k_{\xi}}^\star=Y_{k_{\xi}}.
 \end{cases}
  \end{equation} 
Finally, the waves for the HLLC solver are defined as 
$\mathit{W}^1=U_L^\star-U_L$, $\mathit{W}^2=U_R^\star-U_L^\star$ and $\mathit{W}^3=U_R-U_R^\star$.

\section{Numerical Results}\label{sec:results}
To  assess  the results numerically, we have compared the presented Residual Distribution scheme applied to the model in non-conservative form against the HLLC scheme applied to the model written in the classical conservative form.
In all tests we employ the stiffened gas equations of state.

\subsection{One-dimensional test cases}

All the considered one dimensional tests are Riemann problems, with the initial condition defined by a discontinuity separating two constant states. All the tests are performed with and without mass transfer. The  numerical results with no mass transfer are compared with the exact Riemann problem solution. To calculate this exact solution
we have extended to the four-equation model the methodology presented in \cite{EPkamm,kamm} for the
Euler equations with arbitrary equation of state. Let us note that in these one-dimensional benchmarks not all data points have been represented with a mark to improve line visibility.

\subsubsection{Shock tube test for a liquid-vapor mixture}\label{CBStest}

\vspace{-0.3cm}
The first considered  numerical experiment has been presented in Chiapolino et al. \cite{Chiapolino2017}. 
This test consists in a shock tube  filled with a mixture of liquid and vapour water.
The computational domain is the interval $[0,1]$~m and we set an initial discontinuity at $x=0.5$~m.
Denoting with the subscripts $L$ and $R$ respectively the states on the left and on the right of the
discontinuity, we set the velocities  $u_L=u_R=0$, the temperatures  $T_L=394.2489$~K and $T_R=372.8827$~K, and the pressures  $P_L=2\cdot10^5$~Pa and $P_R=1\cdot10^5$~Pa.
The initial mass fractions and the fluid parameters are reported in Table \ref{cbs_eos}.



\begin{table}[h]
\centering
\begin{tabular}{llllllcc}
\specialrule{.1em}{.05em}{.05em} 
Phase & Fluid & $Y_1$ & $c_v \;[\frac{J}{kg\, K}]$ & $\gamma$ & $P_{\infty}\;[Pa]$ & $q\;[\frac{J}{kg}]$ & $q'\;[\frac{J}{kg\, K}]$\\
\specialrule{.05em}{.05em}{.05em} 
1 & Liquid & $0.2$ & $1816$ & $2.35$ & $1\cdot 10^{9} $ & $-1167\cdot 10^3$ & $0$\\
2 & Gas & $0.8$ & $1040$ & $1.43$ & $0$ & $2030\cdot 10^3$ & $-23.4\cdot 10^3$\\
\specialrule{.1em}{.05em}{.05em} 
\end{tabular}
\caption{Initial mass fractions  and fluid parameters of the water liquid and vapour phases for the  problem of Section \ref{CBStest}.}\label{cbs_eos}
\end{table}

\begin{figure}[H]
\centering
\includegraphics[width=0.45\textwidth]{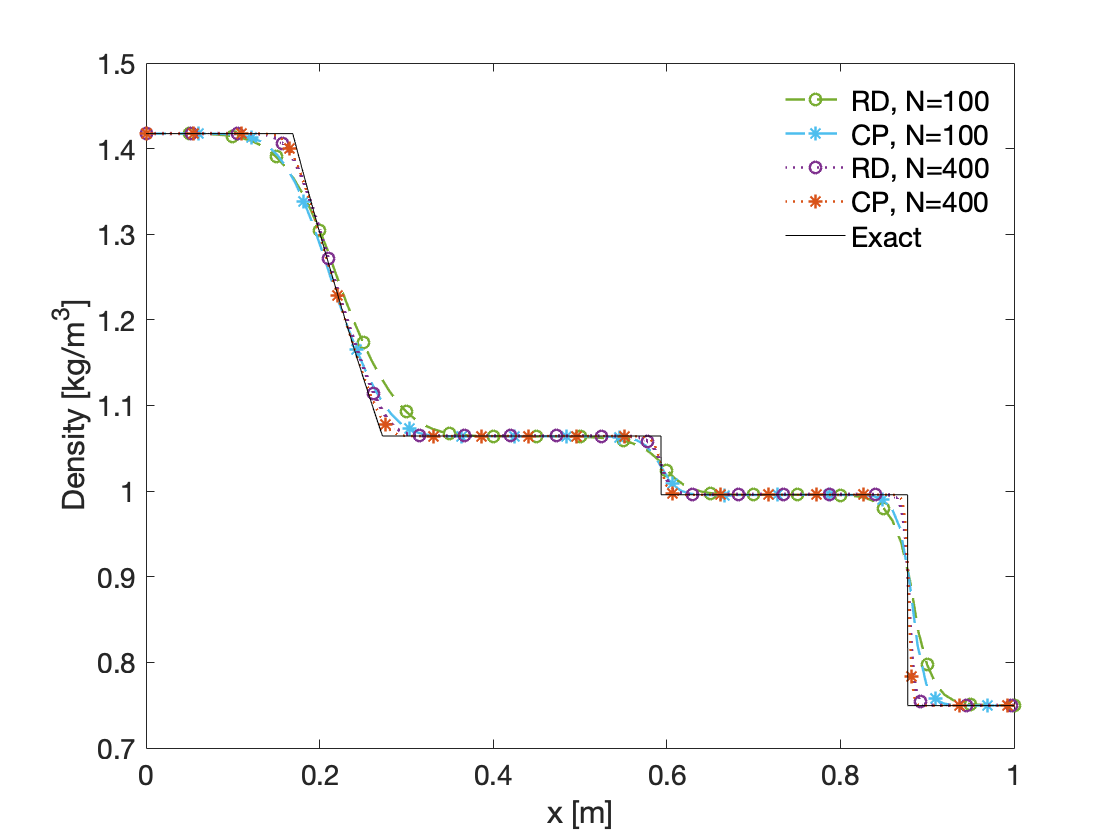}
\includegraphics[width=0.45\textwidth]{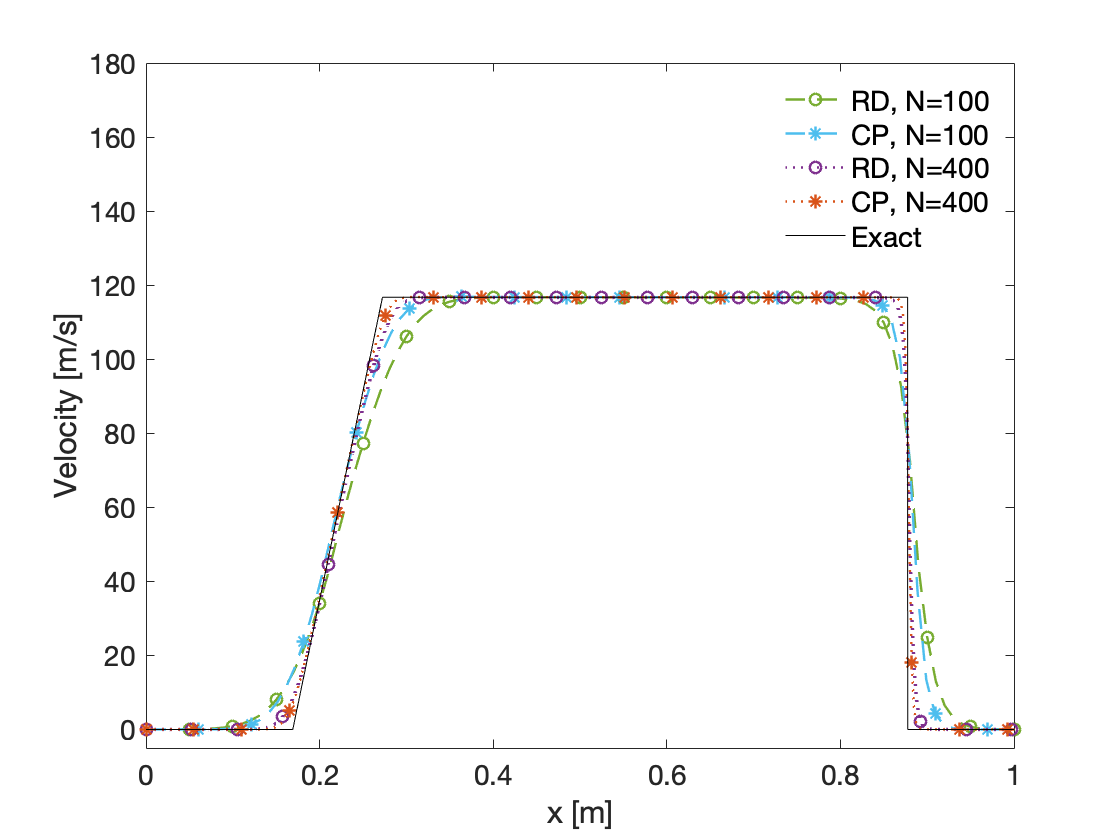}\\
\includegraphics[width=0.45\textwidth]{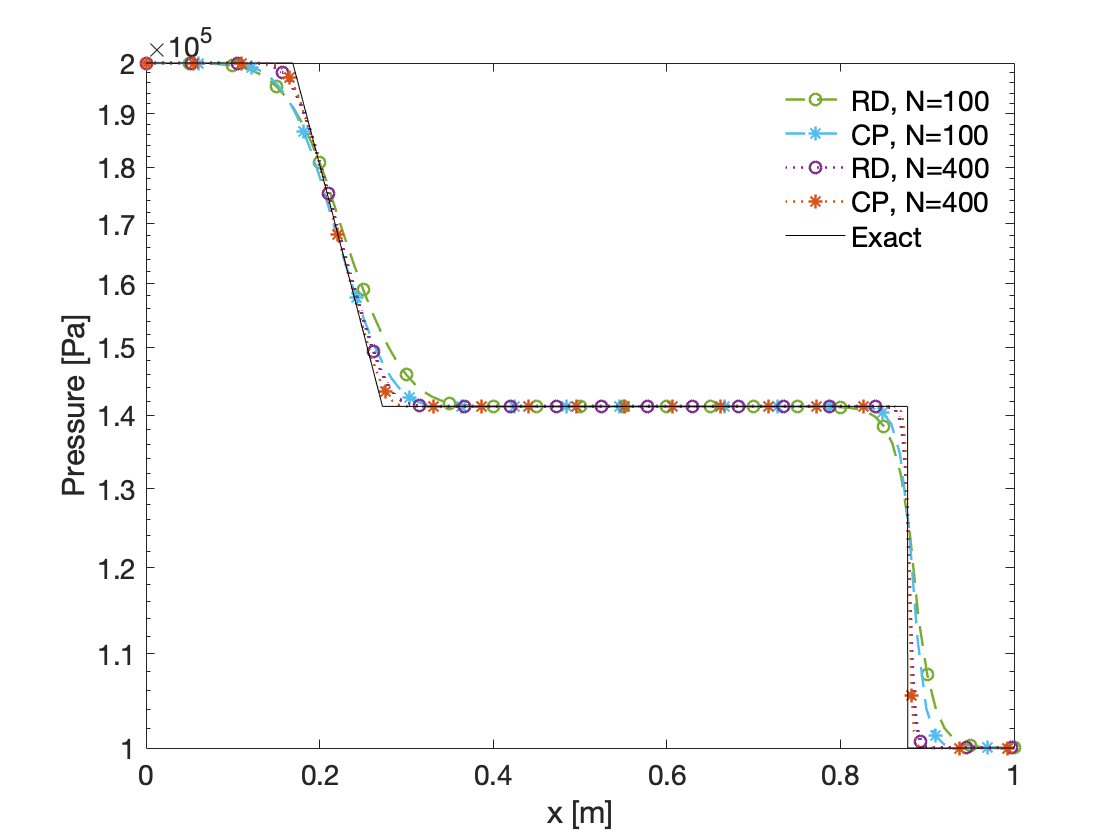}
\includegraphics[width=0.45\textwidth]{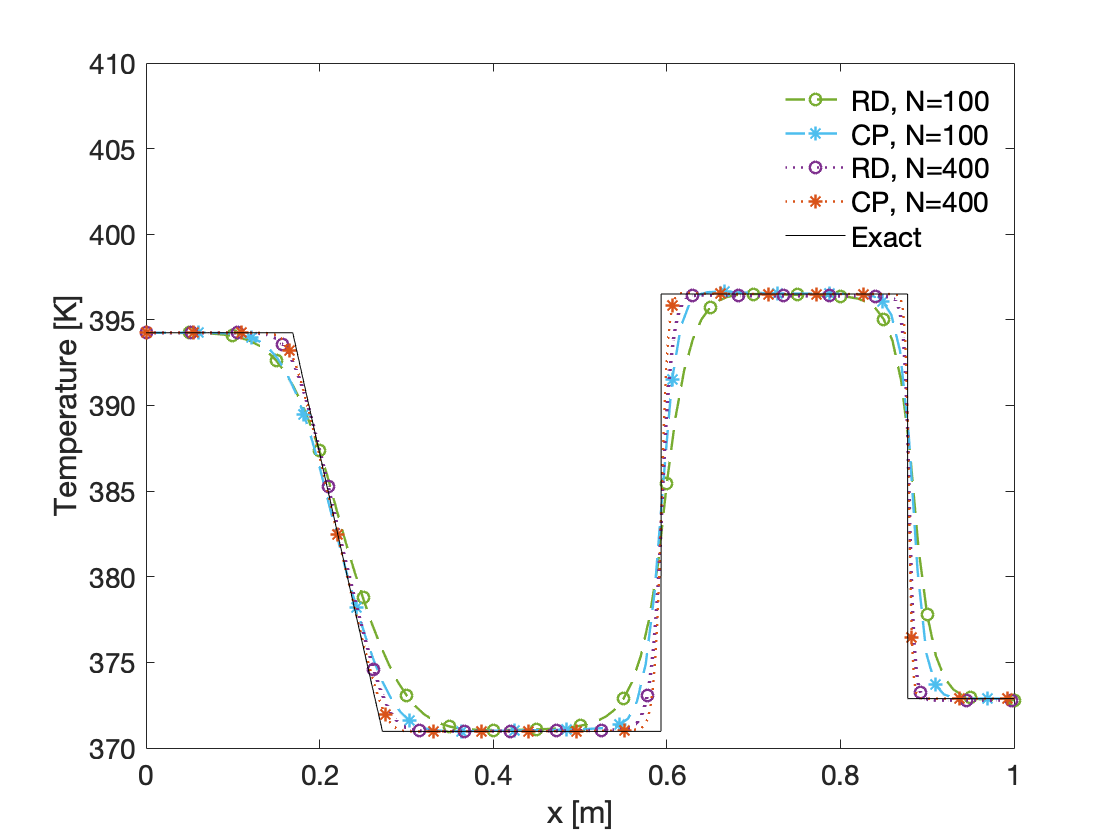}
\caption{Shock tube test for a water liquid-vapour mixture at $t=0.8$~ms. Comparison between the RD and CP approach for the density, velocity, pressure and temperature on different meshes \textit{without} mass transfer.}\label{cbs_nomass}
\end{figure}

\begin{figure}[H]
\centering
\includegraphics[width=0.45\textwidth]{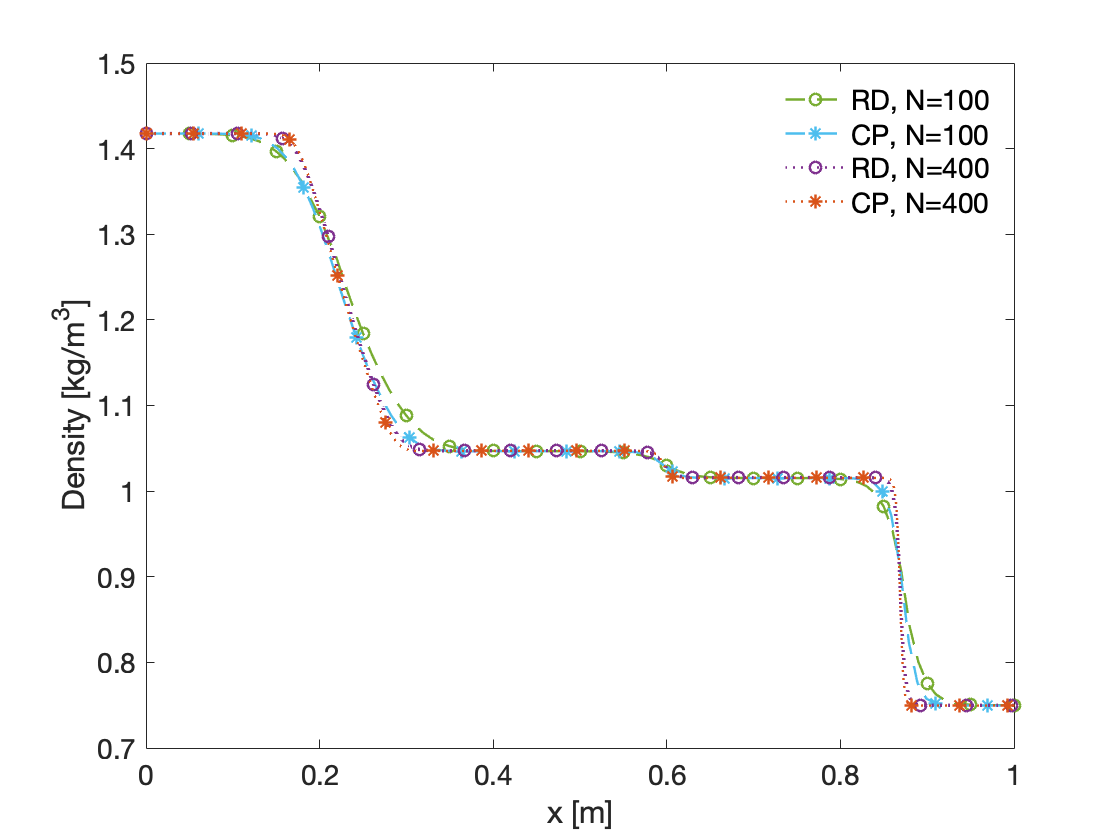}
\includegraphics[width=0.45\textwidth]{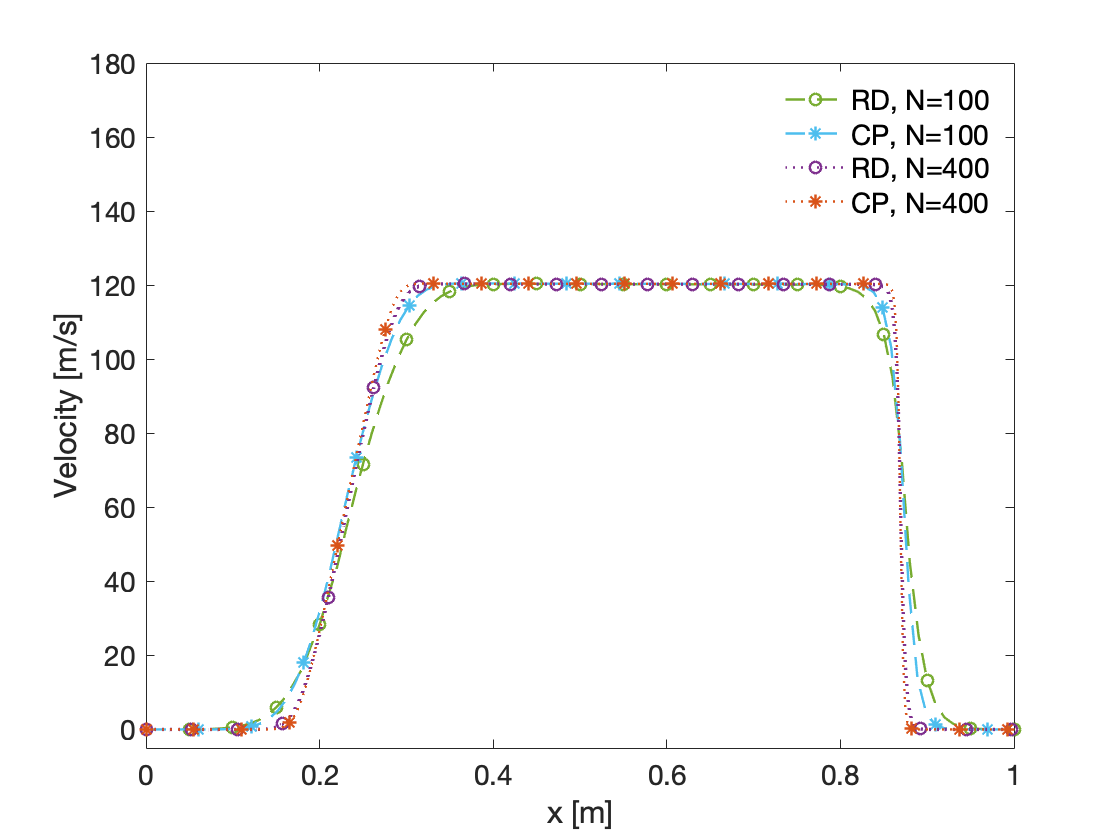}\\
\includegraphics[width=0.45\textwidth]{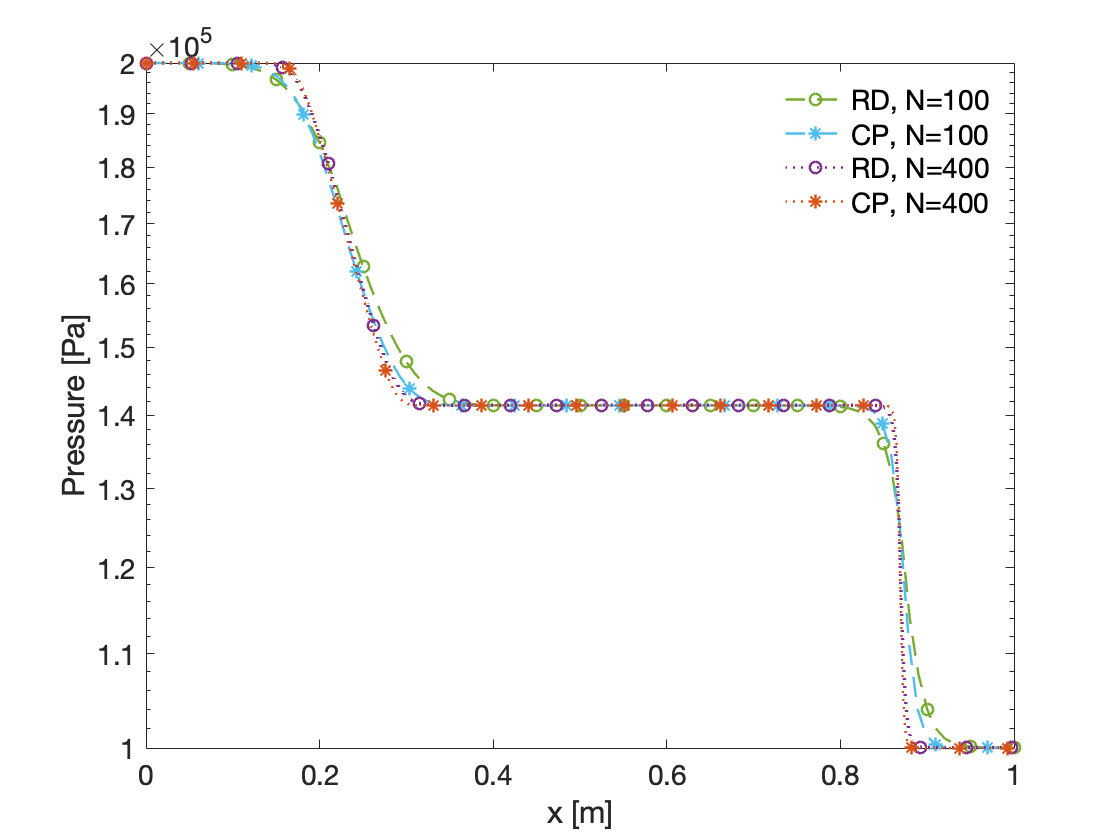}
\includegraphics[width=0.45\textwidth]{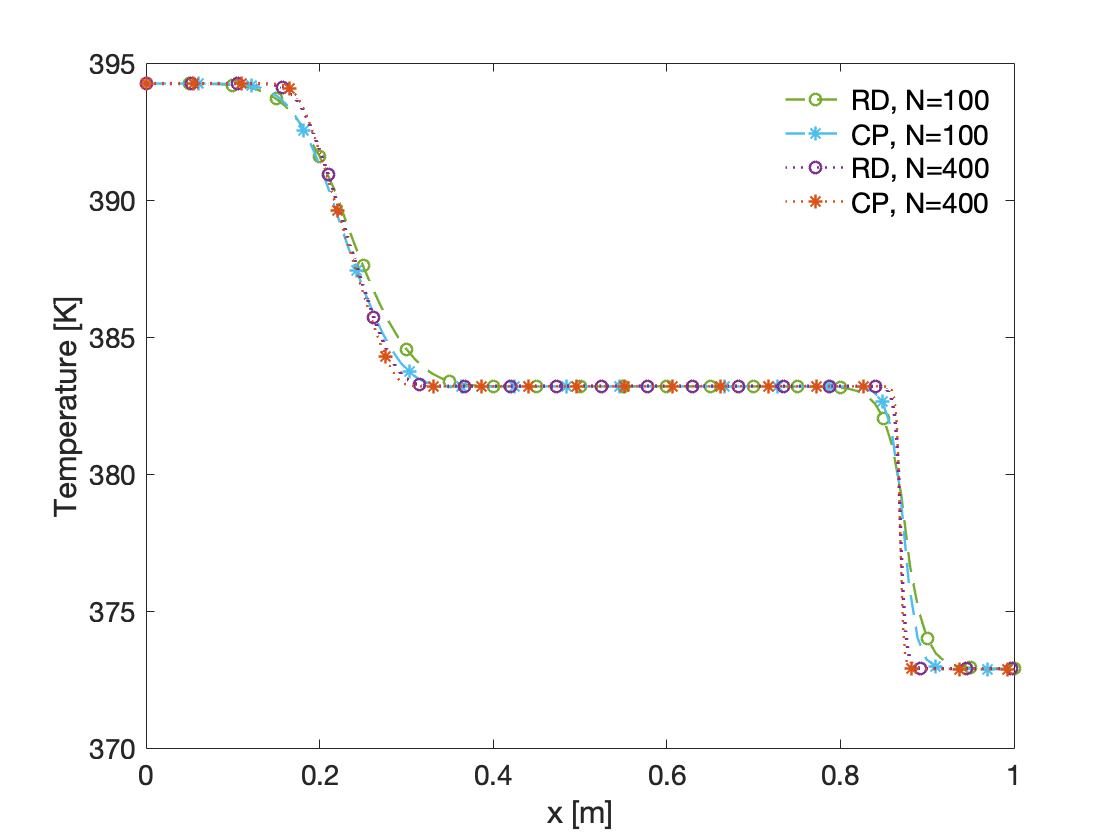}
\caption{Shock tube test for a water liquid-vapour mixture at $t=0.8$~ms. Comparison between the RD and CP approach for the density, velocity, pressure and temperature  on different meshes \textit{with} mass transfer.}\label{cbs_mass}
\end{figure}

\vspace{-0.4cm}
We have compared the non-conservative RD method to  the conservative one based on \textsc{Clawpack} (CP)  and the results are shown in Figures \ref{cbs_nomass}-\ref{cbs_mass}.  In particular,  Figure \ref{cbs_nomass} displays the results for the density,  velocity, pressure and temperature for the  problem \eqref{4_equation_convective} without mass transfer, and these results are also compared against the exact solution.
We note that the RD approximation produces more dissipation with respect to the CP one,  nevertheless it has the advantage of
not requiring a Riemann solver.  Overall, the solution calculated by both methods agrees with the exact solution and as the mesh is refined, the solution converges toward the expected one.
In Figure~\ref{cbs_mass} we show the numerical approximations to system~\eqref{4_equation_source}, which includes mass transfer.
Again one may notice the convergence of the solution as the mesh is refined and there is a good agreement between the non-conservative RD approximation and the conservative CP one.

\subsubsection{Depressurization of a pipe with $CO_2$}\label{Lundtest}
\begin{figure}[H]
\centering
   \includegraphics[width=0.45\textwidth]{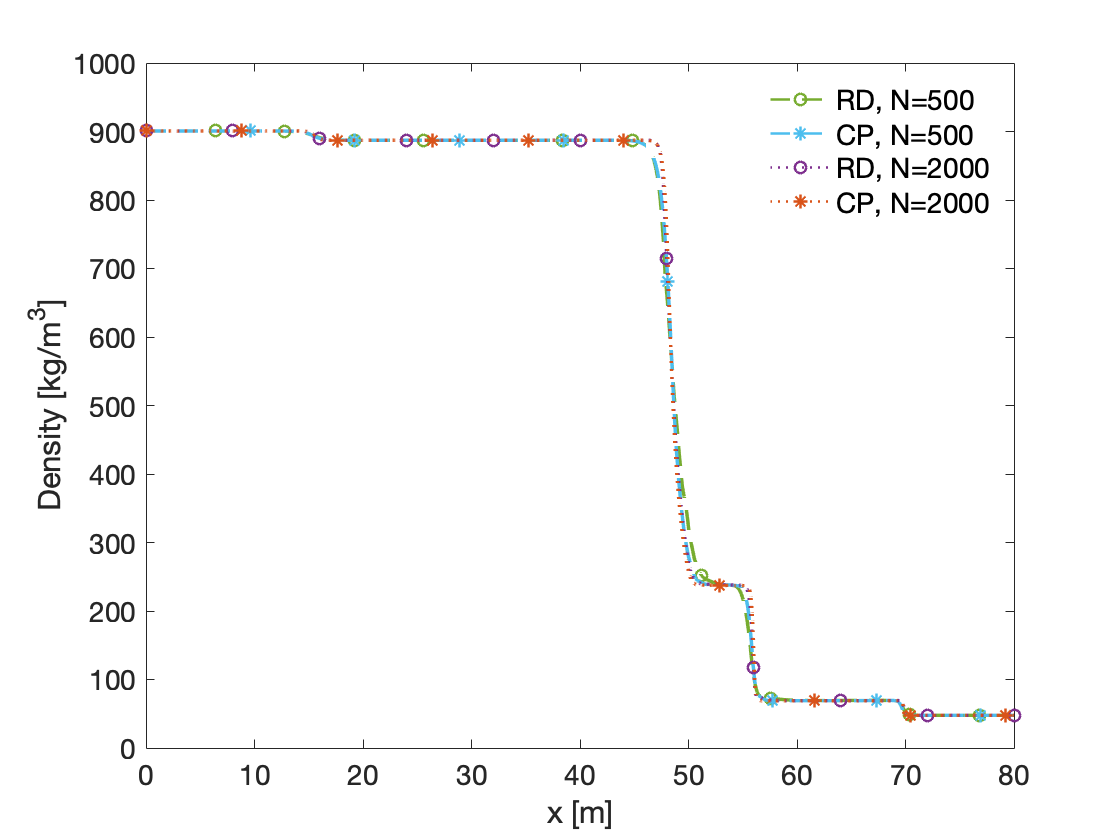}
    \includegraphics[width=0.45\textwidth]{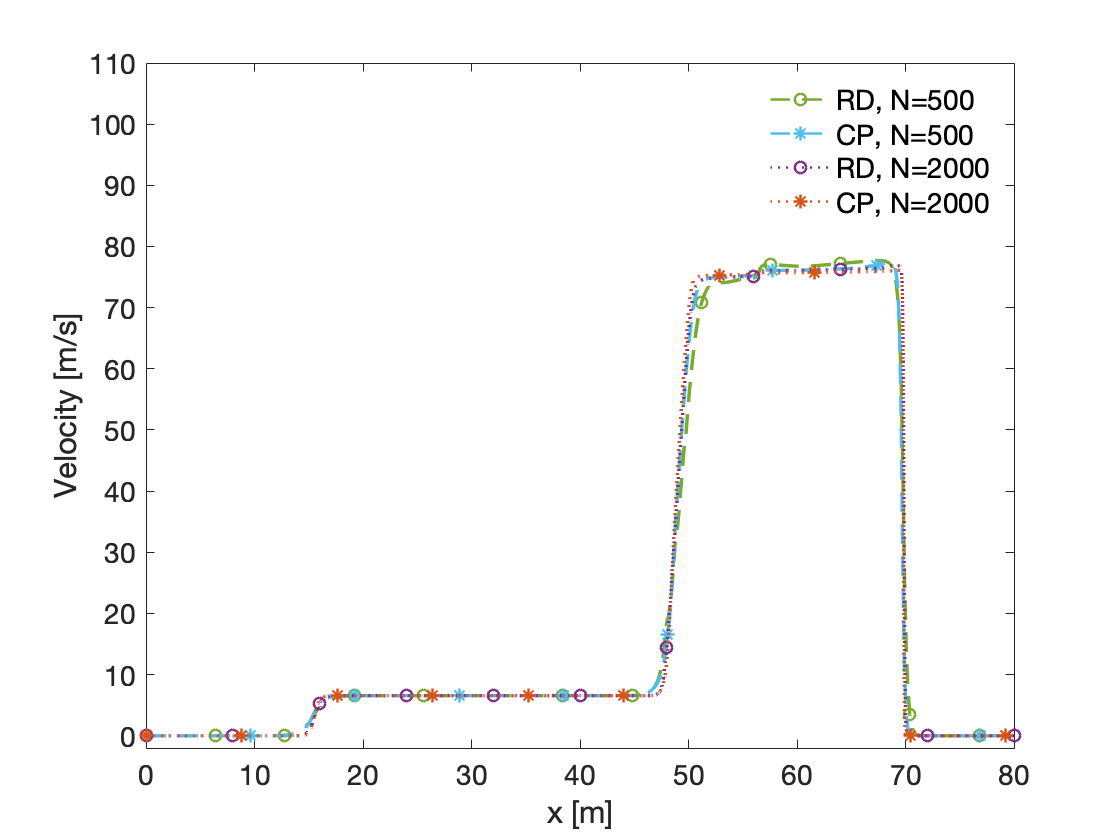}\\
     \includegraphics[width=0.45\textwidth]{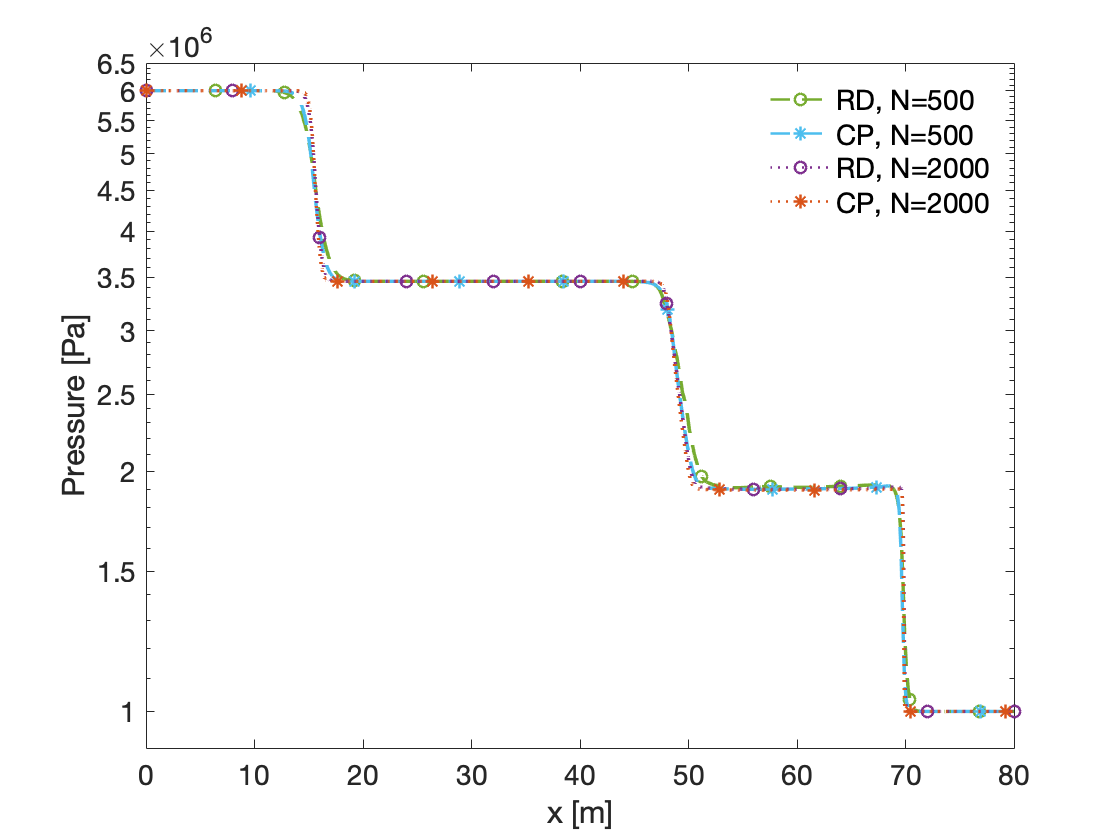} 
    \includegraphics[width=0.45\textwidth]{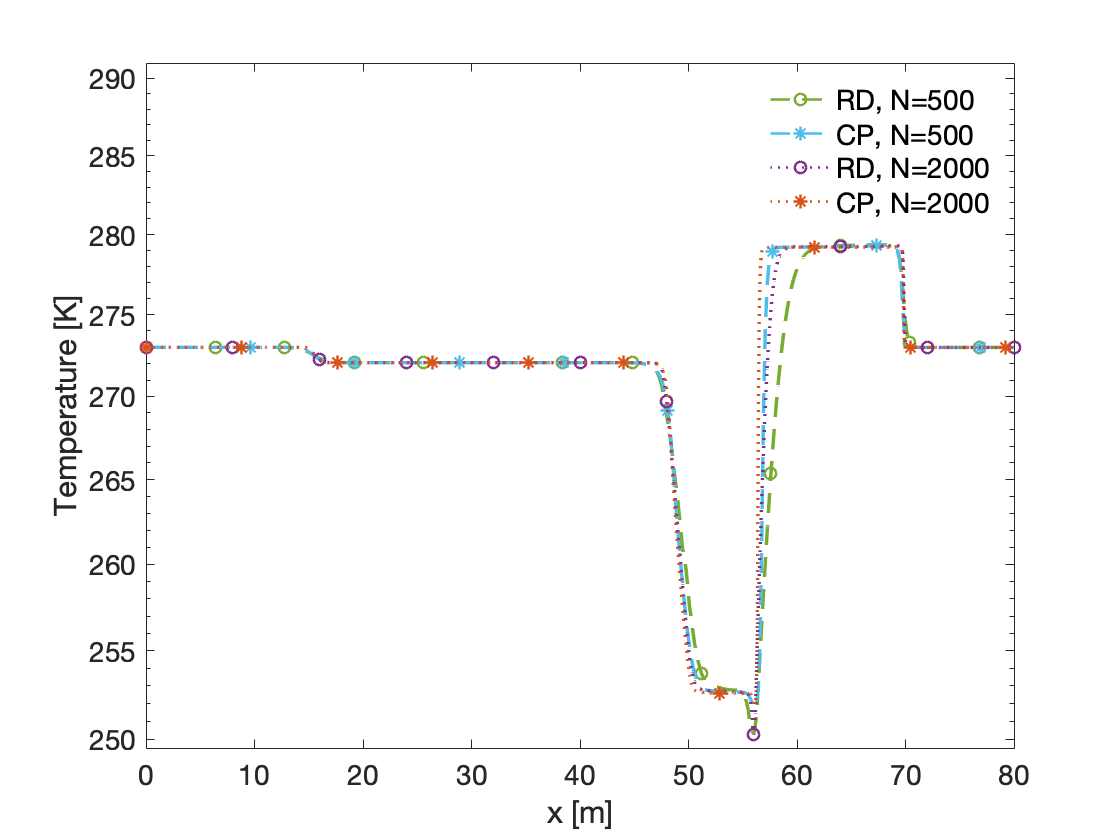} 
     \caption{Depressurisation of a pipe with pure $CO_2$ at $t=0.08$s.  Comparison between the RD and CP approximation for the density, velocity, pressure and temperature on different meshes \textit{with} mass transfer.}\label{CO2_mass}
\end{figure}

\vspace{-0.3cm}
$ \qquad $ As a further test problem, we have considered the depressurization of a pipe 
 filled with $CO_2$, following the benchmark problem proposed in \cite{LundAursand2012}. The total length of the pipe is $L=80$~m. The computational
domain is $[0,80]$~m and we set an initial discontinuity at $x=50$~m. The initial conditions on the left and right of $x$ are in terms of pressures $P_L=60\cdot 10^5$~Pa and $P_R=10\cdot 10^5$~Pa, with an initial velocity $u_L=u_R=0$~m/s, a temperature $T_L=T_R=273$~K and a volume fraction for the liquid phase $\alpha_L=0.99999$ and for the gaseous one $\alpha_R=10^{-5}$.
The equation of state parameters are shown in Table \ref{co2_eos}.
Since the original reference paper of this benchmark problem does not explicitly write all the considered parameters, we have chosen some parameters with freedom, as the volume fraction and the parameter $q'$, as reported in Table \ref{co2_eos}.
\begin{table}[H]
\centering
\begin{tabular}{lllllcc}
\specialrule{.1em}{.05em}{.05em} 
Phase & Fluid  & $c_v \;[\frac{J}{kg\, K}]$ & $\gamma$ & $P_{\infty}\;[Pa]$ & $q\;[\frac{J}{kg}]$ & $q'\;[\frac{J}{kg\, K}]$\\
\specialrule{.05em}{.05em}{.05em} 
1 & Liquid  & $2.44\cdot10^3$ & $1.23$ & $1.32\cdot 10^{8} $ & $-6.23\cdot 10^5$ & $-5.3409289\cdot10^3$\\
2 & Gas & $2.41\cdot10^3$ & $1.06$ & $8.86\cdot 10^{5}$ & $-3.01\cdot 10^5$ & $-1.0398090\cdot 10^4$\\
\specialrule{.1em}{.05em}{.05em} 
\end{tabular}
\caption{Equation of state parameters for the $CO_2$ tube test in Section \ref{Lundtest}.}\label{co2_eos}
\end{table}
\begin{figure}[H]
\centering
\includegraphics[width=0.45\textwidth]{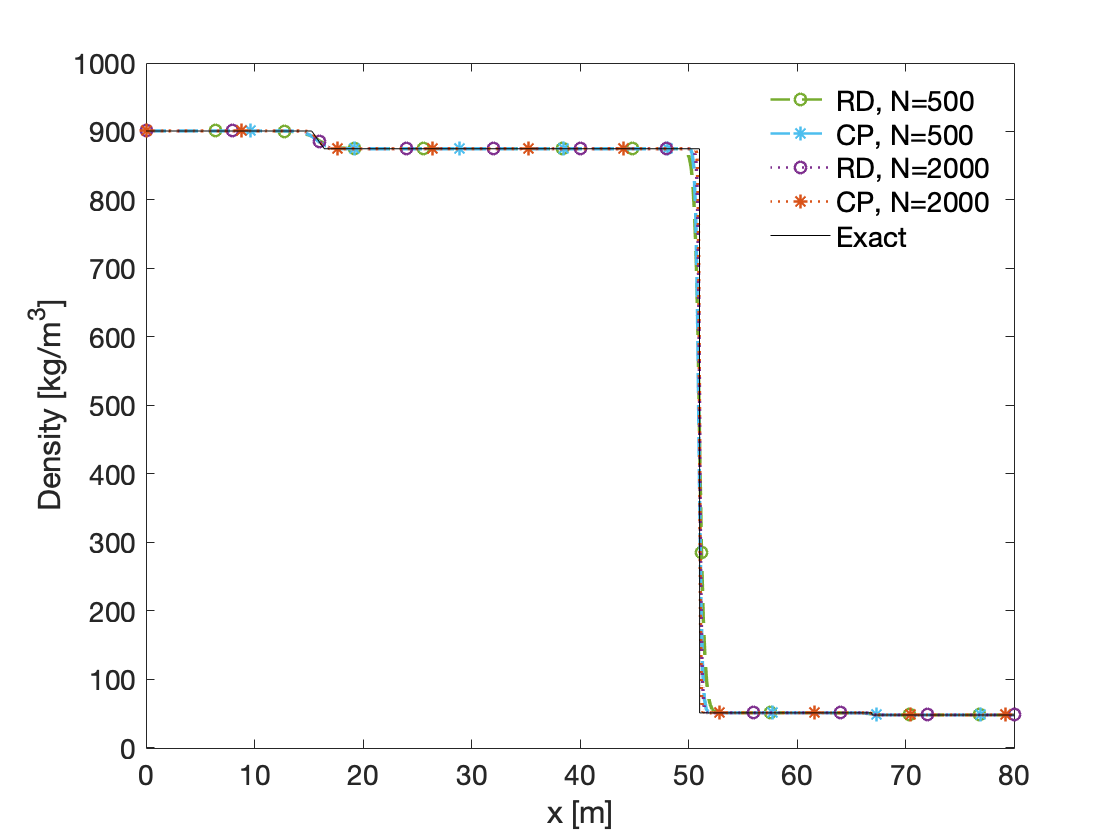}   
 \includegraphics[width=0.45\textwidth]{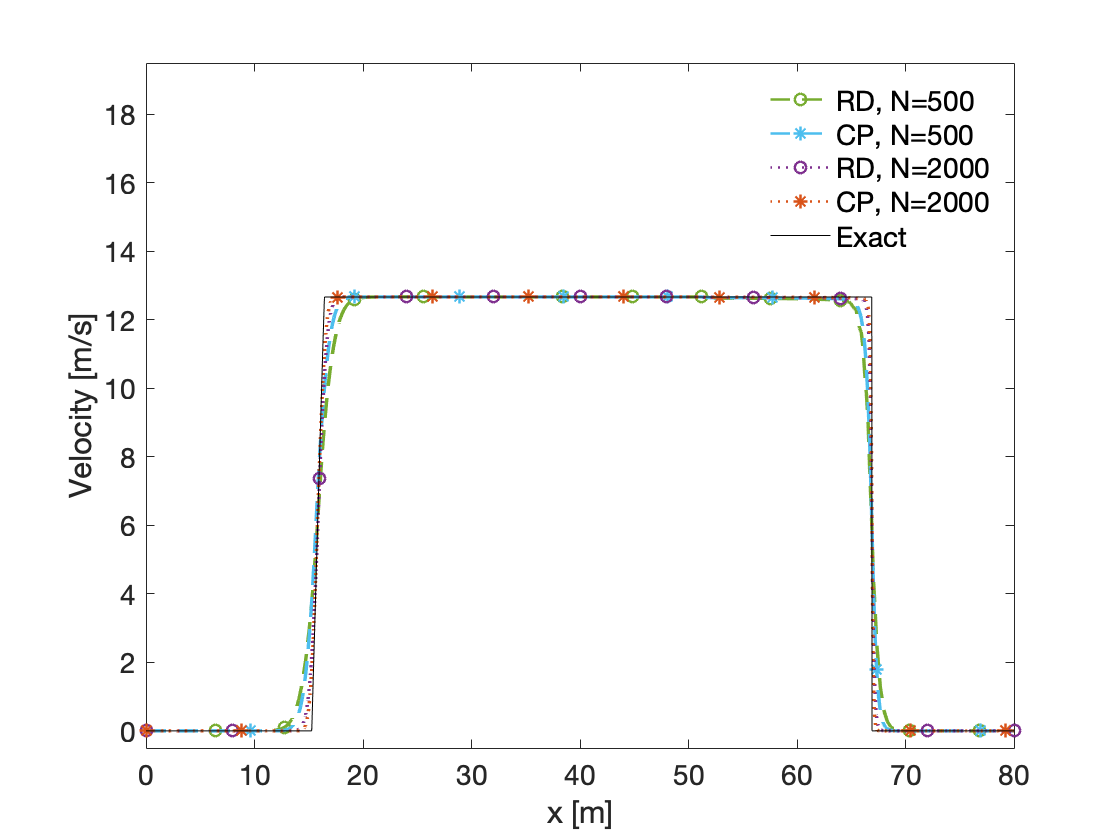}\\
  \includegraphics[width=0.45\textwidth]{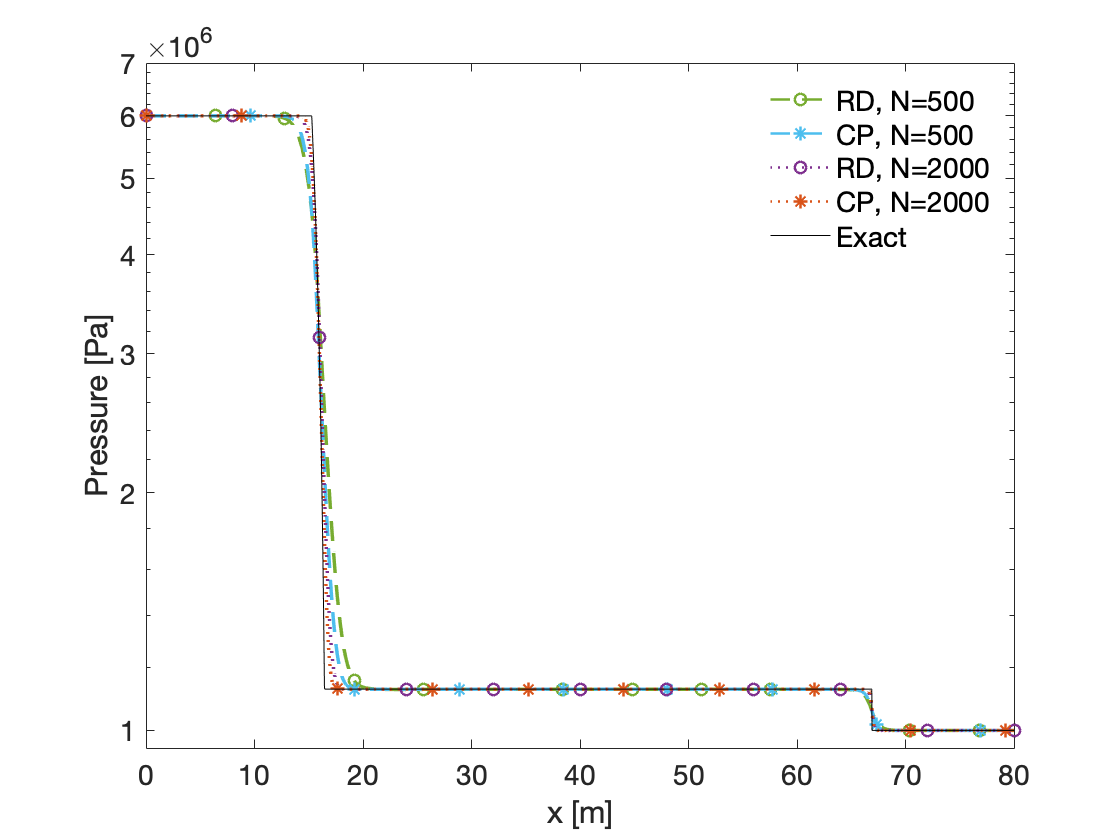} 
  \includegraphics[width=0.45\textwidth]{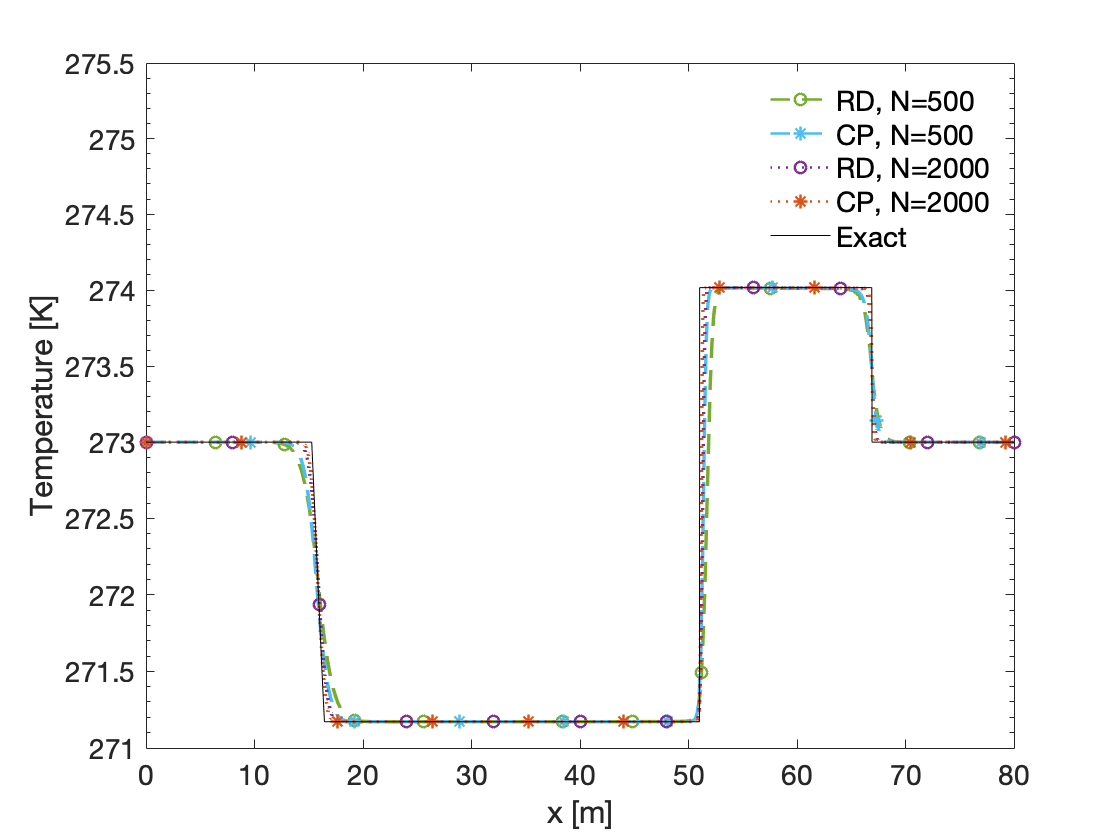}
      \caption{Depressurisation of a pipe with pure $CO_2$ at $t=0.08$s.  Comparison between the RD and CP approximation for the density, velocity, pressure and temperature on different meshes \textit{without} mass transfer.}\label{CO2_nomass}
\end{figure} 
We show in Figures~\ref{CO2_nomass}-\ref{CO2_mass} the results computed by the RD and CP methods at time $t=0.08$~s without and with mass transfer for the density, the velocity, the pressure and the temperature. For the case with no mass transfer we also display the exact solution of the problem.
 We observe a very good agreement of both the non-conservative RD approximation and the CP one with the exact solution for the problem without mass transfer.
For the problem with mass transfer we observe overall a good agreement between the CP and RD solution, as well a good qualitative agreement with the results in \cite{LundAursand2012}.

\subsubsection{Cavitation test for water}\label{cavitation_2}

\begin{figure}[H]
\centering
  \includegraphics[width=0.45\textwidth]{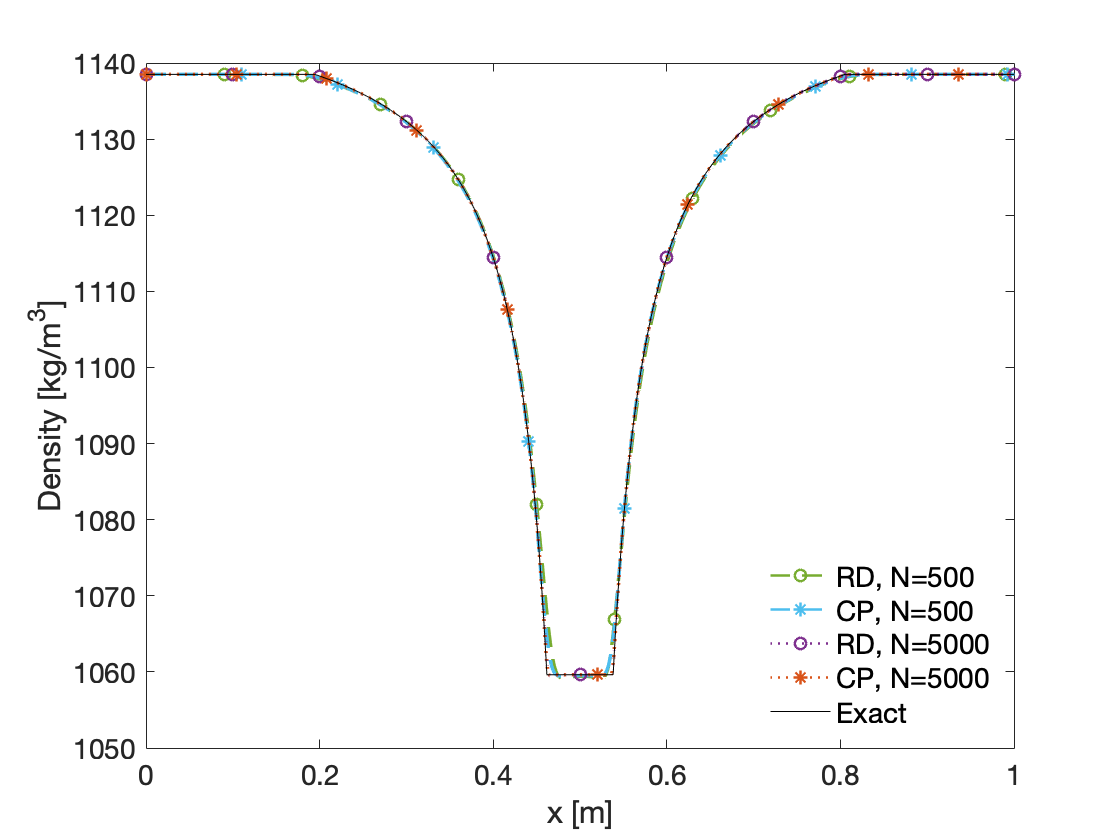}\includegraphics[width=0.45\textwidth]{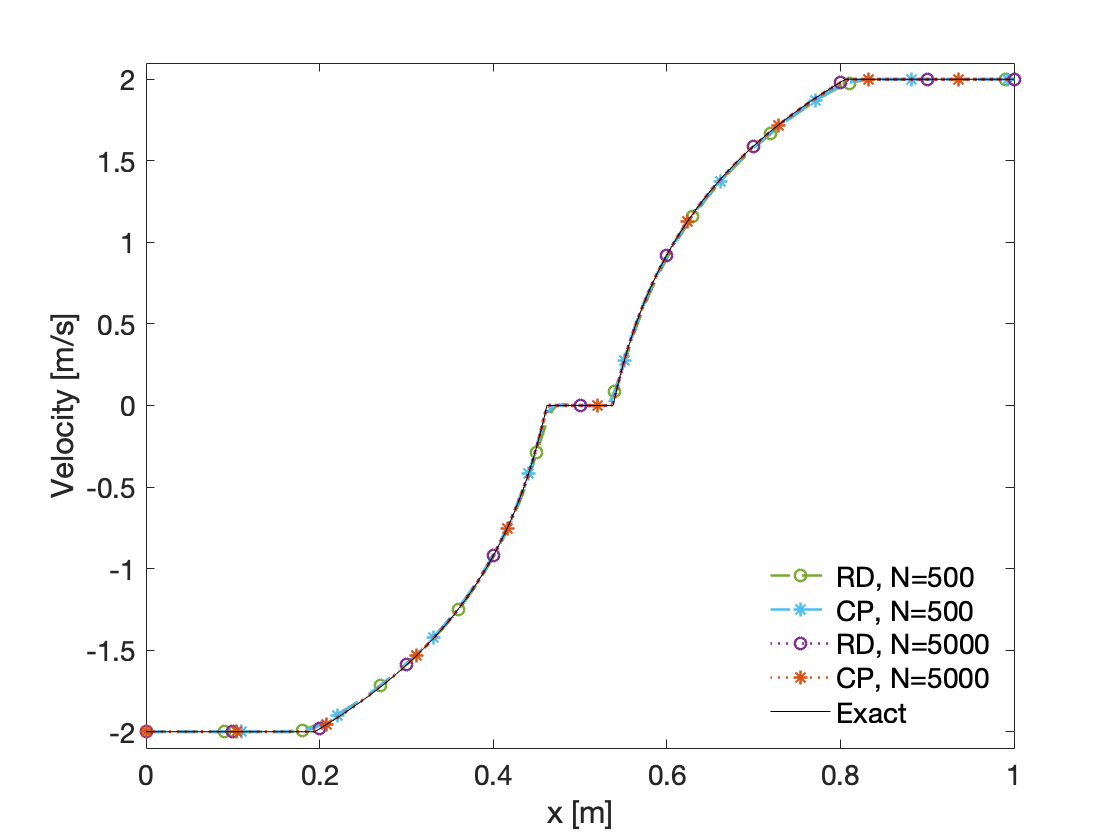}\\
    \includegraphics[width=0.45\textwidth]{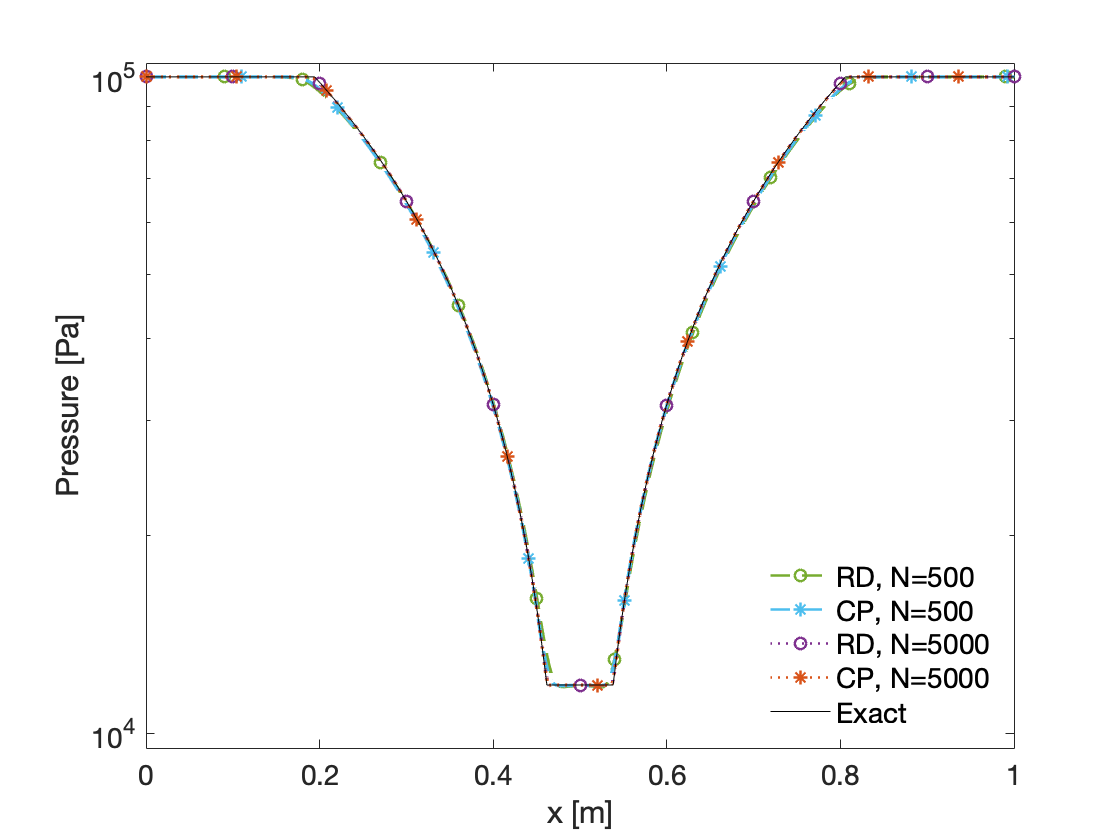}  \includegraphics[width=0.45\textwidth]{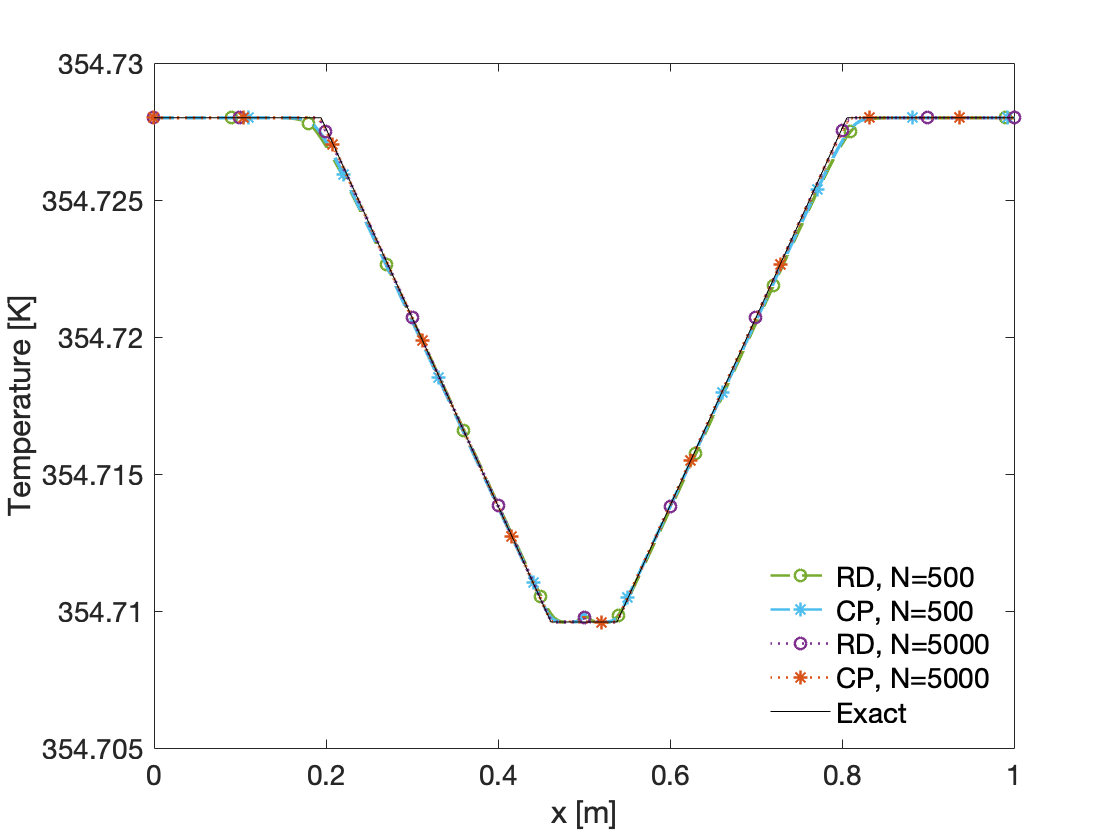}\\
\caption{Water cavitation tube problem at $t=3.2$ms. Comparison between the RD and CP approximation for the density, velocity and temperature on different mesh sizes \textit{without} mass transfer.}\label{cavitation2_nomass}
\end{figure}

\begin{figure}[H]
\centering
  \includegraphics[width=0.45\textwidth]{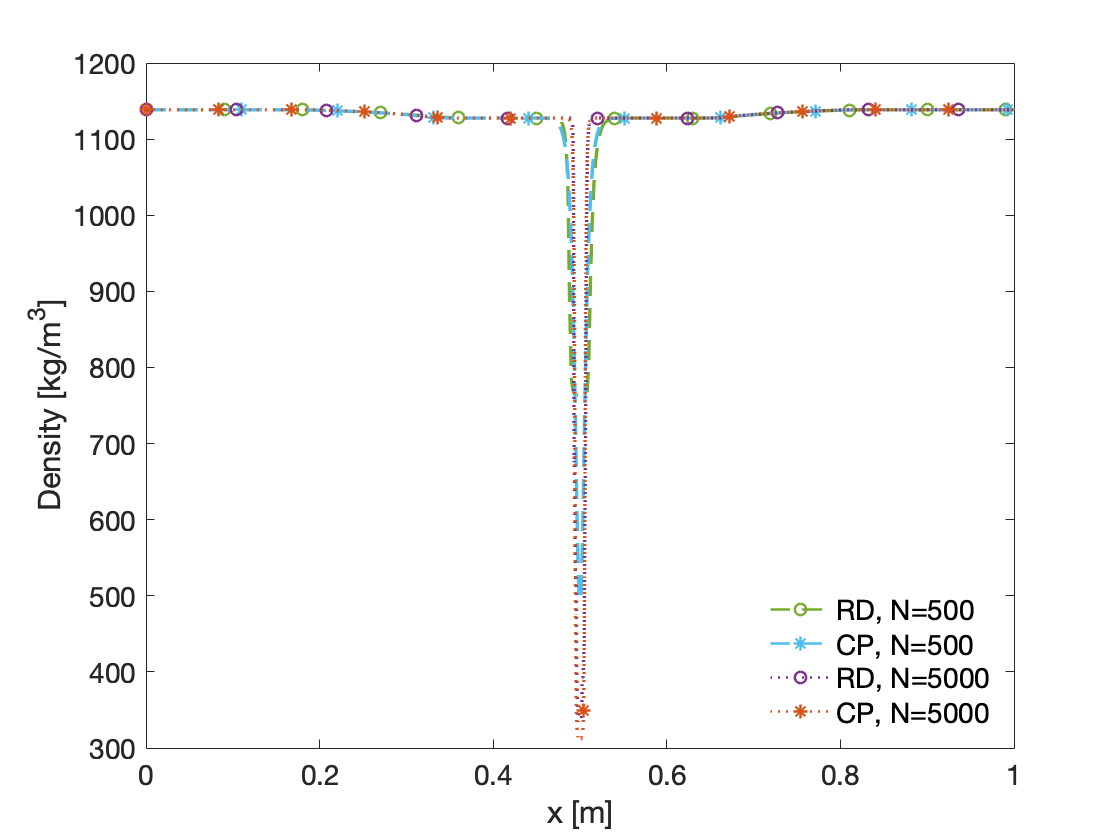}
    \includegraphics[width=0.45\textwidth]{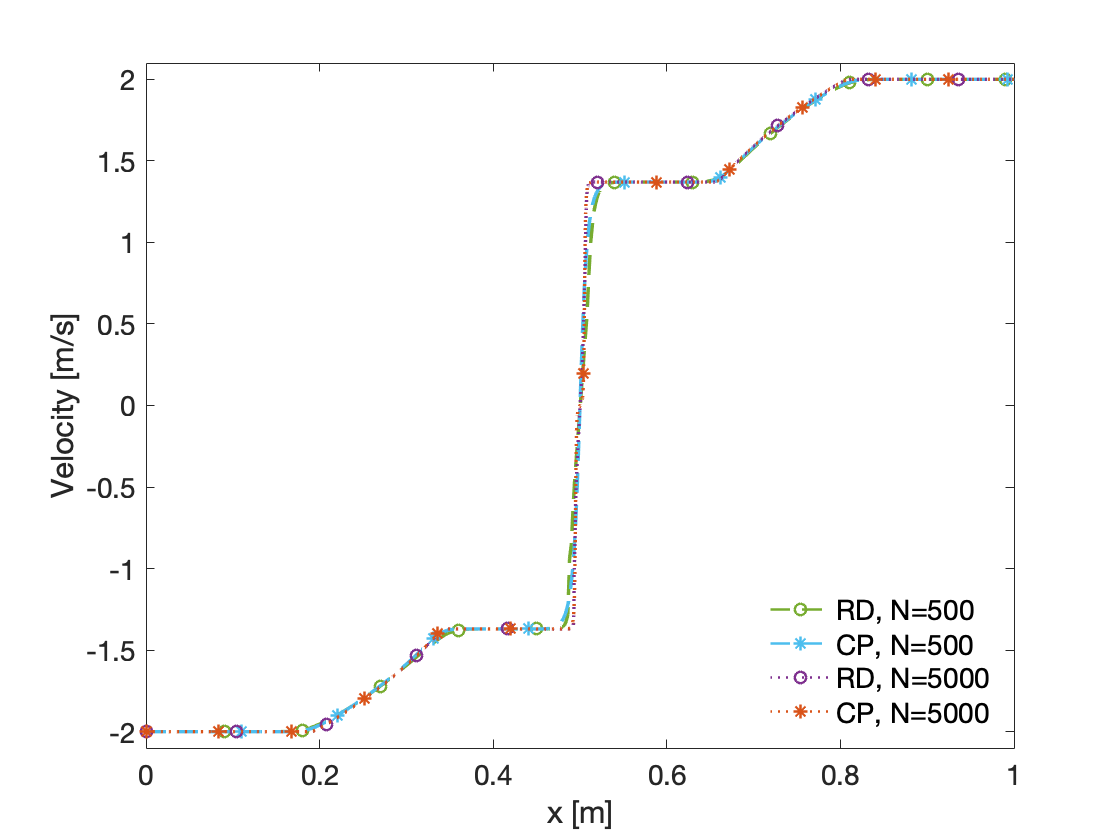}\\
     \includegraphics[width=0.45\textwidth]{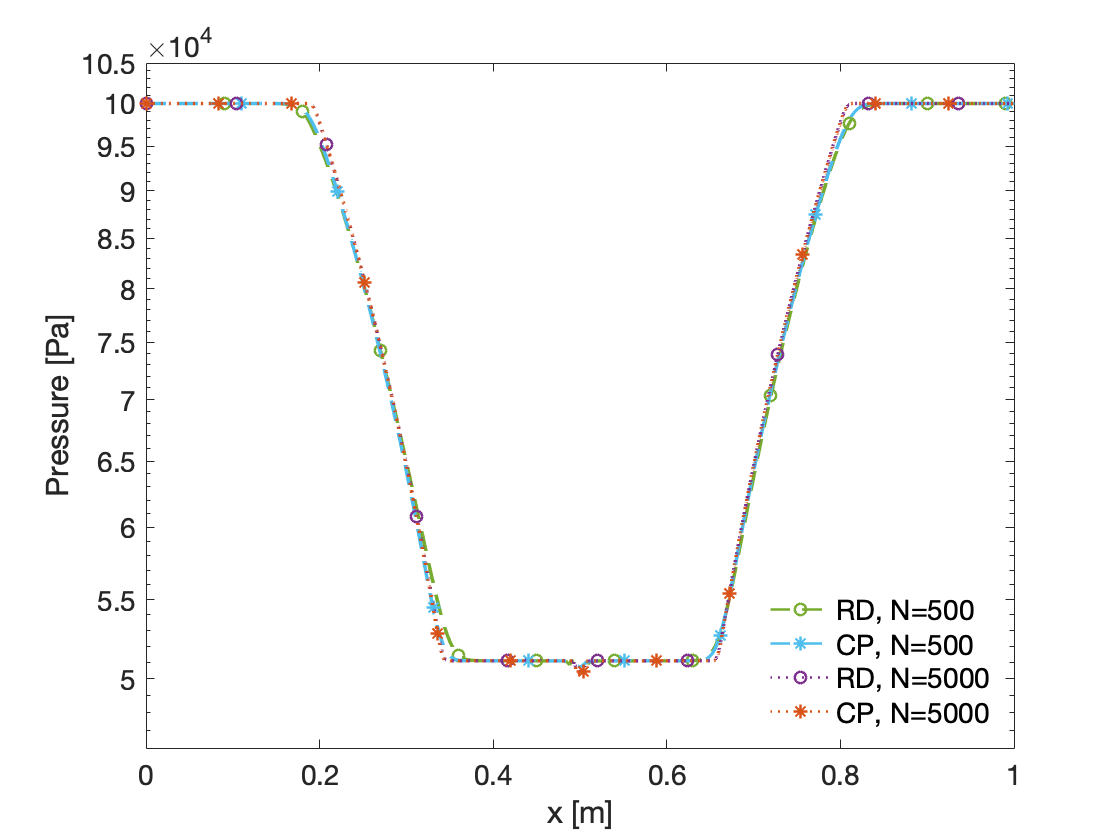}
     \includegraphics[width=0.45\textwidth]{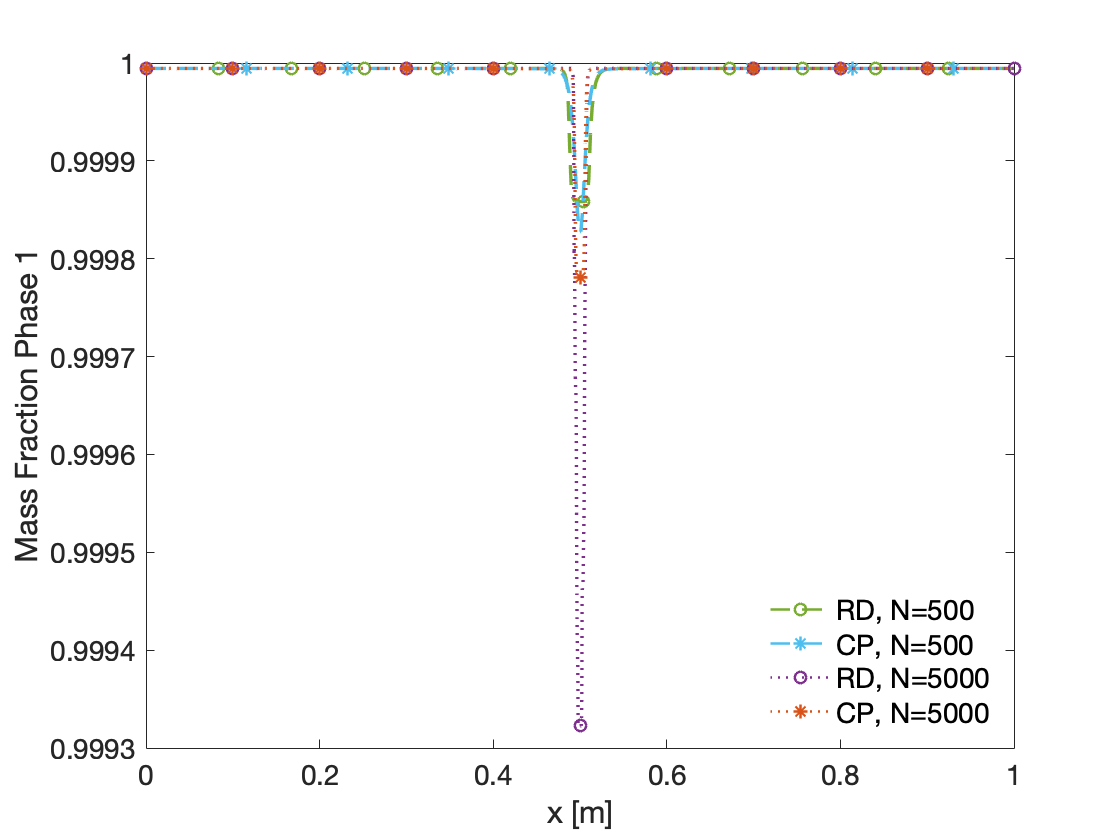}
\caption{Water cavitation tube problem at $t=3.2$ms. Comparison between the RD and CP approximation for the density, velocity and temperature on different mesh sizes \textit{with} mass transfer.}\label{cavitation2_mass}
\end{figure}

\vspace{-0.3cm}
We now consider a classical benchmark problem for Baer--Nunziato type
two-phase models, a cavitation tube test (see e.g.\ \cite{Saurel2008,Pelanti2014}). We set a discontinuity at $x=0.5$~m within an interval $[0,1]$~m. The initial velocity on the left and right of the discontinuity is  $u_L=-2$m/s and $u_R=2$m/s, respectively. Further, we set for the liquid a density $\rho_{liq}=1150$ kg/m$^3$ and for the vapour $\rho_{vap}=0.63$ kg/m$^3$. The liquid volume fraction is set to $\alpha_{vap}=0.01$. The pressure at the initial time is $P=10^5$~Pa and the temperature $T=354.728$~K.
The parameters used for the stiffened gas equations of state are reported in Table~\ref{cbs_eos}.\\
Results obtained by the RD and CP methods without mass transfer
are displayed in Figure~\ref{cavitation2_nomass} and compared to the exact solution of the problem. Very good agreement is observed.
Results obtained with mass transfer are shown in 
Figure~\ref{cavitation2_mass}, and good agreement is noticed between the two considered methods.


\subsection{Two-dimensional test cases}
\subsubsection{Shock helium-bubble interaction without mass transfer over a structured grid}\label{2D_bubble}

\begin{figure}[H]
\centering
\includegraphics[width=0.6\textwidth]{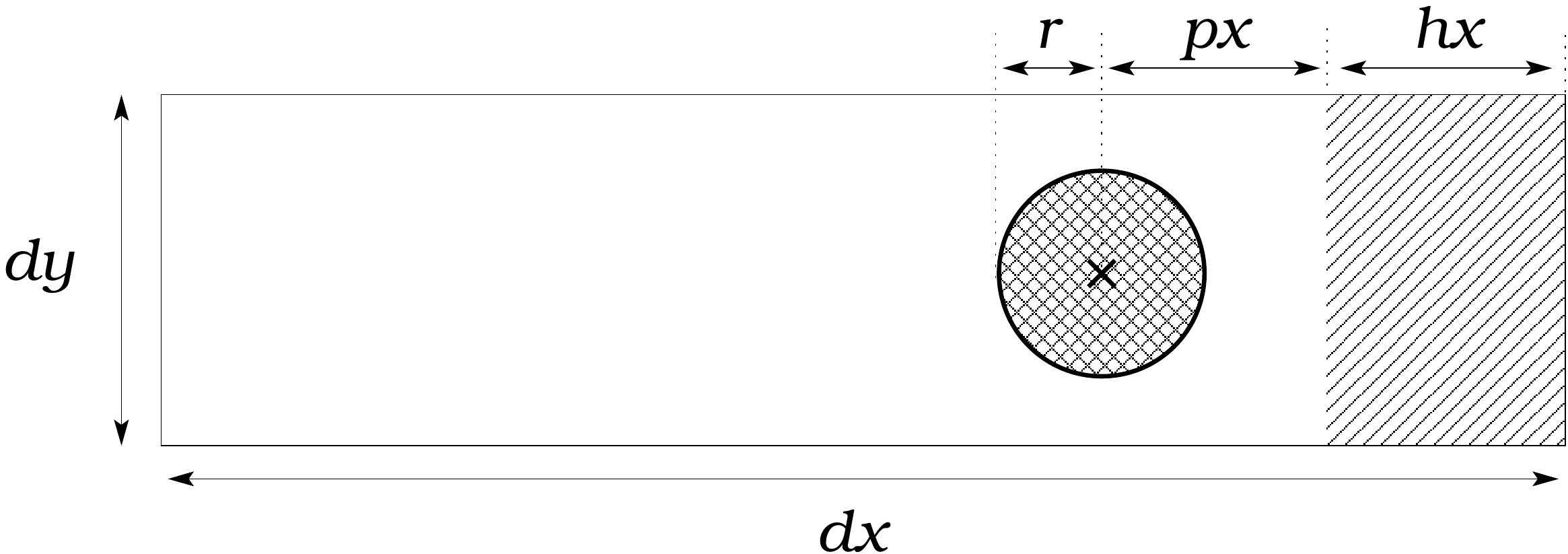}
\caption{2D shock helium-bubble interaction. Sketch of the benchmark problem.
}\label{Fig:Bubble_config}
\end{figure}

\vspace{-0.3cm}
In the following we consider a benchmark in 2D without considering the mass transfer terms. In particular, we have the configuration as detailed in Figure \ref{Fig:Bubble_config}, following the experiments presented in \cite{Layes2007} and the numerical simulation in \cite{Saurel2009}.
Here we simulate a shock propagating in air at Mach 1.5 interacting with a bubble of Helium.
The domain has a length $dx=0.3$~m and a height $dy=0.08$~m and there is an initial diaphragm at a distance $hx=0.015$~m from the right edge that separates two regions of the air with conditions corresponding to pre-shock (left) and post-shock (right) states. The Helium bubble has radius $r=0.02$~m and its center is located at a distance $px=0.027$~m from the diaphragm.
The values of pressure,  density, and velocity in the three regions (left and right regions with respect to the diaphragm, and bubble region) are specified in the Table~\ref{bubble_setup}.
Helium and air are modeled as perfect gases with $\gamma_{He}=1.67$ and $\gamma_{air}=1.4$, respectively. As mentioned above, for numerical reason each  pure fluid region (air and gas) contains a small amount of the other fluid ($\alpha_k =10^{-9}$).
 Outflow conditions are set on the left and right domain boundaries, while wall conditions are set on the lower and upper bounds along the x-axis length.

\begin{table}[H]
\centering
\begin{tabular}{lccc}
\specialrule{.1em}{.05em}{.05em} 
  Area & $\rho  \;{\rm [\frac{kg}{m^3}]}$ &$u\;\left[\frac{m}{s}\right]$&$P \; [\rm Pa]$\\
\specialrule{.05em}{.05em}{.05em} 
 Left  & $1.29$  & $0$ & $1.01325\cdot 10^5$\\
 Bubble & $0.1669$& $0$ &$1.01325\cdot 10^5$\\
 Right & $2.4$& $-230.3$ & $2.4909 \cdot 10^5$\\
\specialrule{.1em}{.05em}{.05em} 
\end{tabular}
\caption{Set-up for the different phases in the shock helium-bubble interaction test in Section \ref{2D_bubble}.}\label{bubble_setup}
\end{table}

\vspace{-0.5cm}
\begin{figure}[H]
\centering
\includegraphics[trim={0 0 0 8.5cm},clip,width=6cm,height=3cm]{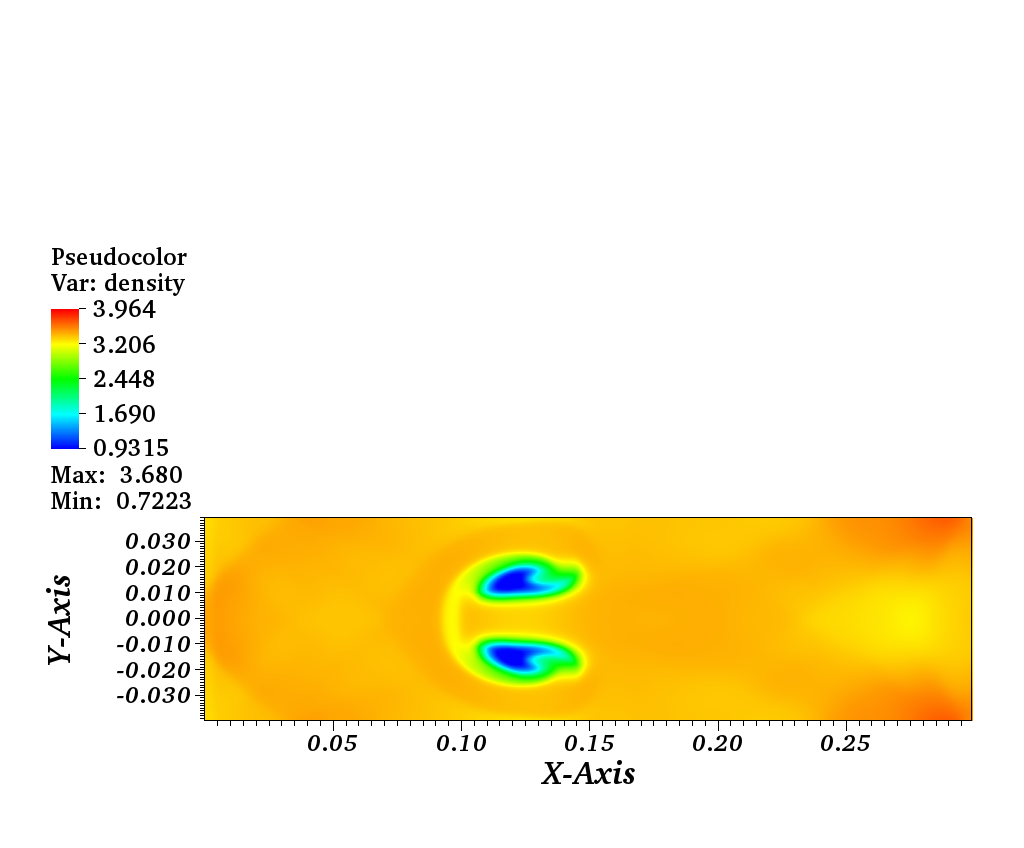}\includegraphics[trim={ 0 0 0 8.5cm},clip,width=6cm,height=3cm]{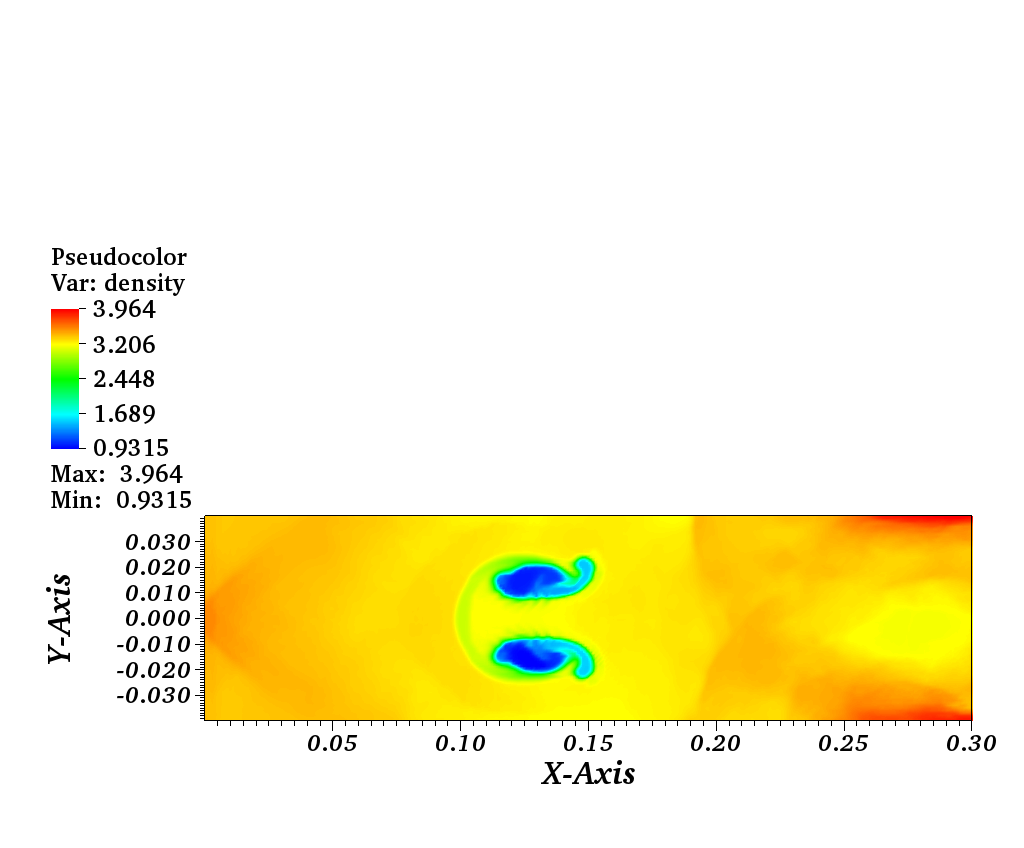}
\caption{2D shock helium-bubble interaction at $t=577 \mu s$. Density $\rm [kg/m^3]$. Left: CP solver, conservative formulation of the system;
right: RD method, non-conservative formulation of the system.}\label{Shock2D_density}
\end{figure}

\begin{figure}[H]
\centering
\includegraphics[trim={0 0 0 8.5cm},clip,width=6cm,height=3cm]{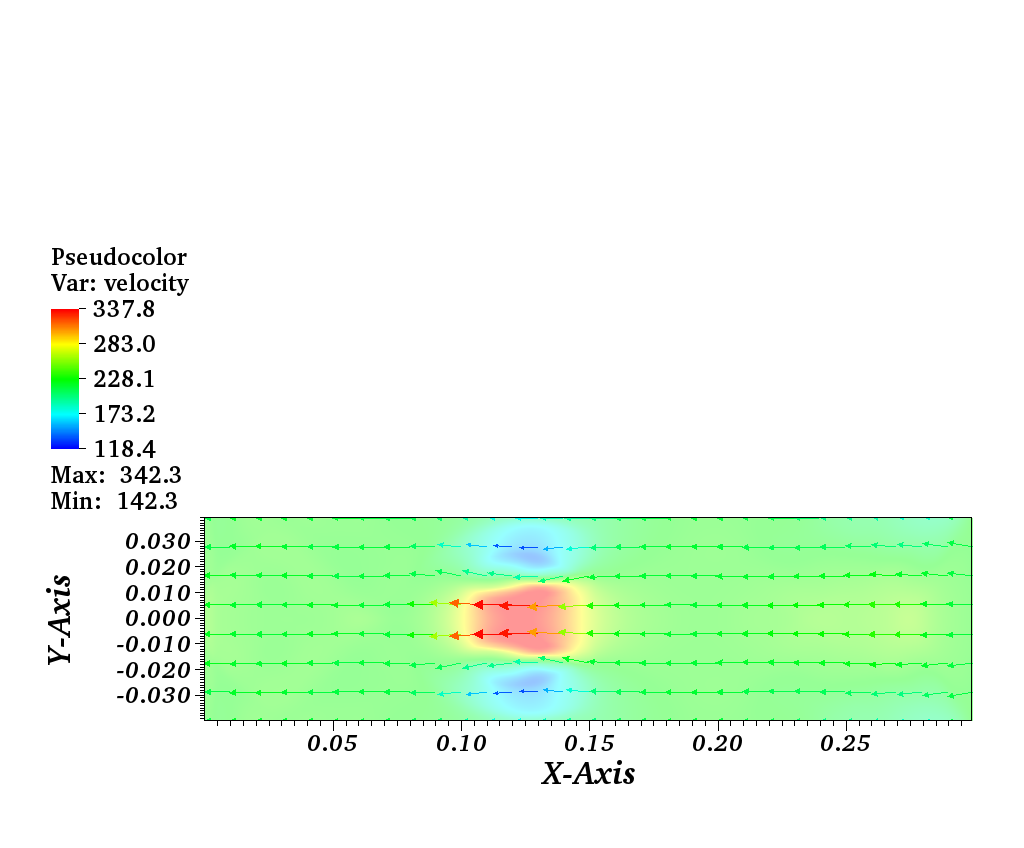}\includegraphics[trim={ 0 0 0 8.5cm},clip,width=6cm,height=3cm]{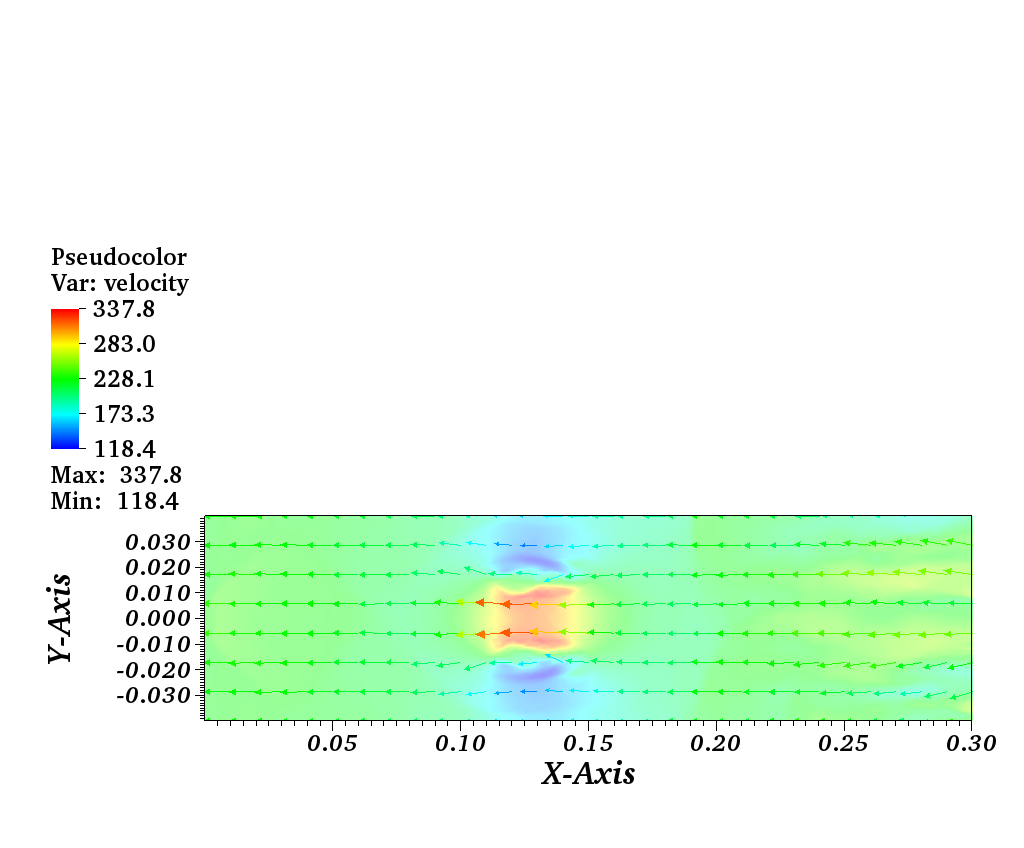}
\caption{2D shock helium-bubble interaction at $t=577 \mu s$. Velocity $\rm [m/s]$. Left: CP solver, conservative formulation of the system;
right: RD method, non-conservative formulation of the system.
}\label{Shock2D_velocity}
\end{figure}

We consider a structured cartesian, uniform mesh, with $N_1=24377$ elements, corresponding  approximately to a grid with $300\times 80$ cells. 
In Figure \ref{Shock2D_density} we display the total density at $t=577 \mu s$. In particular, we compare in the left figure the approximation obtained using CP on a conservative formulation, i.e. \eqref{4eqsmodel_mass_c}, while in the right figure the presented RD strategy on a non-conservative set of equations, \eqref{4eqsmodel_mass_nc}. The right hand side of the system has been neglected in this test, comparing only the solution to the homogeneous hyperbolic part of the system of the four-equation model. The RD formulation is computed considering a CFL=0.1 along with a stabilization term $\theta=0.01$ (c.f. Subsection \ref{Sec:MOOD}).
Figure \ref{Shock2D_density} shows in the area $0.09\,{\rm m}<x<0.16\,{\rm m}$ two flow structures along the symmetry axis in the flow direction created by the velocity air jet through the helium bubble. In particular, as also described in \cite{Layes2007}, we observe how the generated vorticity leads to a downstream and an upstream vortex. While propagating, the difference in velocities of the two vortex rings allows a growth of downstream ring and a reduction of the upstream one. 
Both the CP and RD approximation strategies cross-validate this behaviour.  The RD approach shows an overall more accurate representation of the bubble transformation, w.r.t. the CP one. Nevertheless, it is possible to observe a
slight asymmetry for the RD approximation, most probably linked to the boundary conditions. In Figure \ref{Shock2D_velocity} we represent the velocity along with its vector field for both the CP (left) and RD (right) approximations. Also for this visualization, the RD approach shows a more neat and  detailed bubble approximation, while appears less symmetric due to boundary related perturbations on the solution given at the outflow boundary and which affects the area for $x>0.2$. This could be in part due to the choice of modeling the outflow conditions as subsonic following \cite{desantis2013}. Currently, further approaches are being studied to overcome such asymmetries.

\vspace{-0.5cm}
\subsubsection{Liquid-vapor water mixture with mass transfer over an unstructured grid}\label{2D_CBS}

\begin{figure}[H]
\centering
\subfigure[Sketch set-up]{\includegraphics[width=0.3\textwidth]{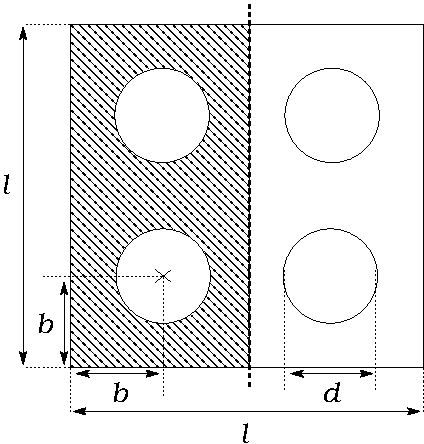}}
\subfigure[Mesh detail]{\includegraphics[width=0.35\textwidth]{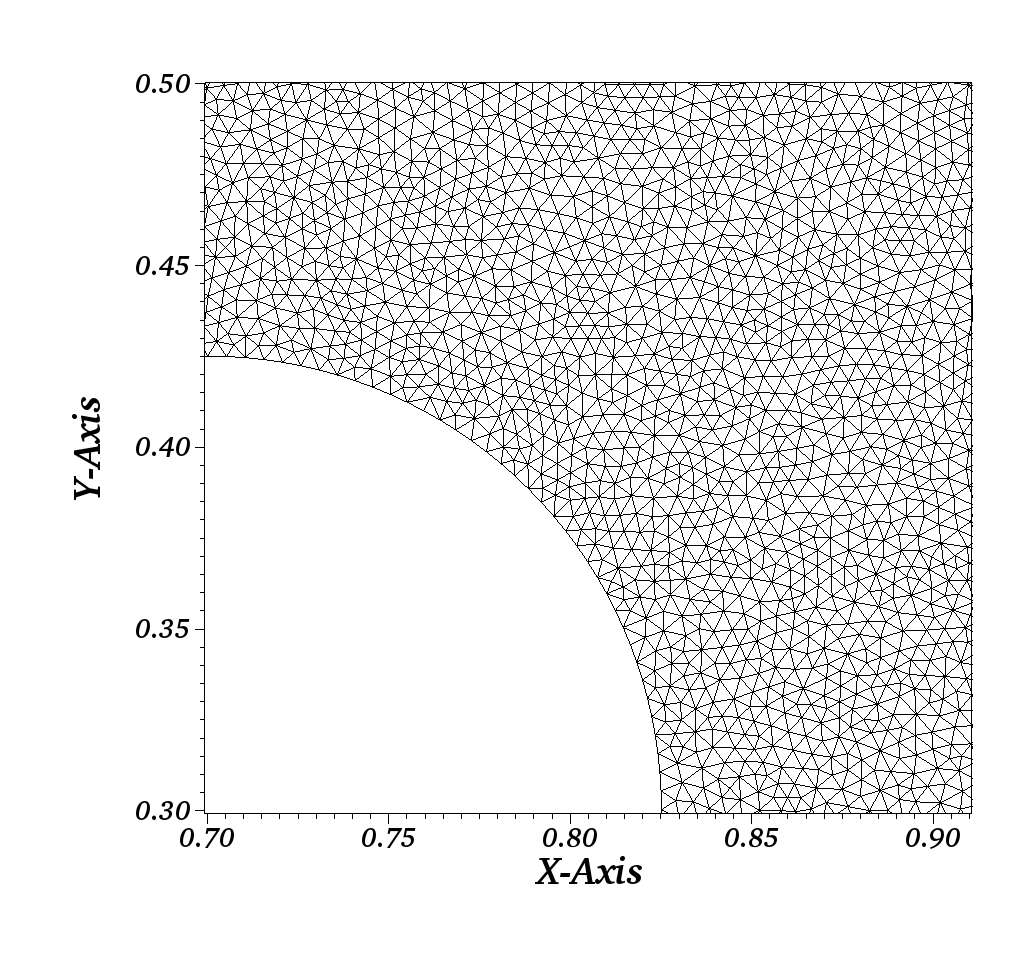}\label{2D_CBS_mesh}}
\caption{2D liquid-vapor water mixture. Sketch of the benchmark problem and zoom around a cylinder quarter to show the mesh type.
}\label{Fig:CBS_config}
\end{figure}

\vspace{-0.3cm}
Finally, we present a test-case dealing with a more complex geometry on an unstructured grid. Specifically, we consider  a square domain of side length $l=1$~m containing four rigid fixed cylinders. We impose on the right and on the left domain boundaries outflow conditions, while on the upper and lower bounds, as well as along the cylinders, wall boundary conditions. The set-up consists in liquid and vapor water in a mixed state 
with the same initial conditions with a discontinuity used for the one-dimensional shock tube test in Section~\ref{CBStest} and with fluid parameters indicated in Table~\ref{cbs_eos}. The initial diaphragm that separates the regions with $P_L=10^5$~Pa and $P_R=2\cdot 10^2$~Pa is set at $x=0.5$~m.
This benchmark design mimics a typical industrial set-up using heat exchangers. 
The goal of such a test is, for instance, to evaluate the effect of a sudden pressure gradient causing phase changes in the flow that can result in very damaging consequences for the cylinders. The considered square domain of $l=1$~m is discretized via an unstructured mesh with $79512$ elements.
The center of each cylinder is set at $b=0.3$~m along the $x$-axis, measured from the nearest outflow boundary and at $b$ in $y$-direction, measured from the nearest wall. Each cylinder has a diameter of $d=0.25$~m. A sketch of the set-up and a zoom on the mesh are shown in Figure \ref{Fig:CBS_config}. 
\begin{figure}[H]
\centering
\subfigure[Density]{\includegraphics[width=0.495\textwidth]{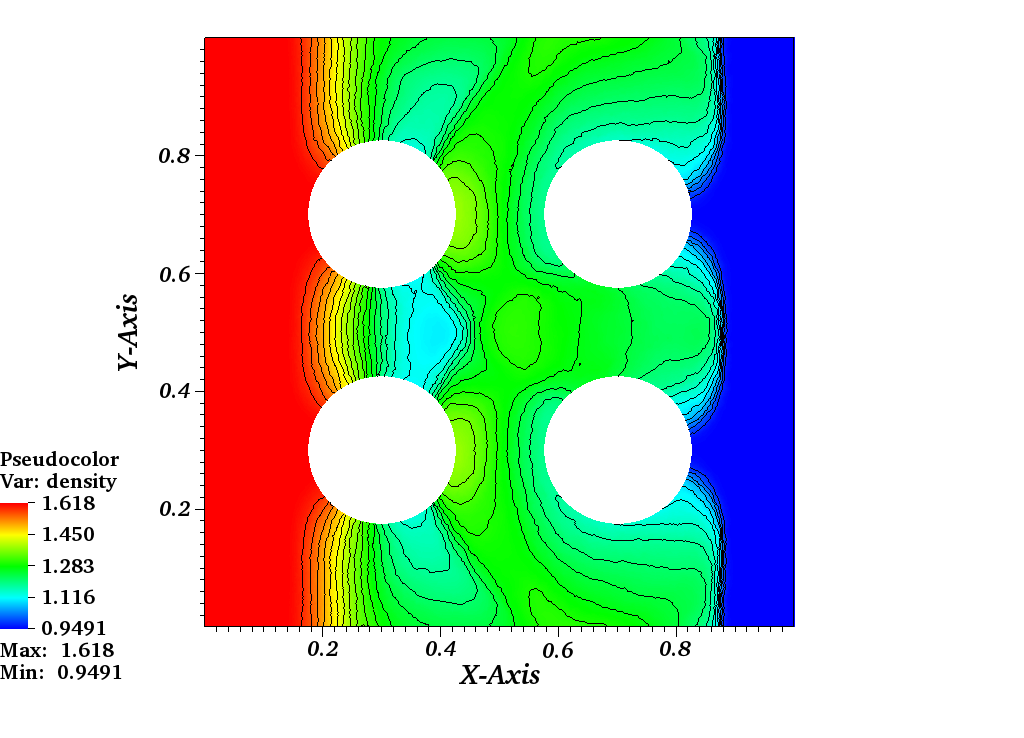}\label{2D_CBS_density}}
\subfigure[Pressure]{\includegraphics[width=0.495\textwidth]{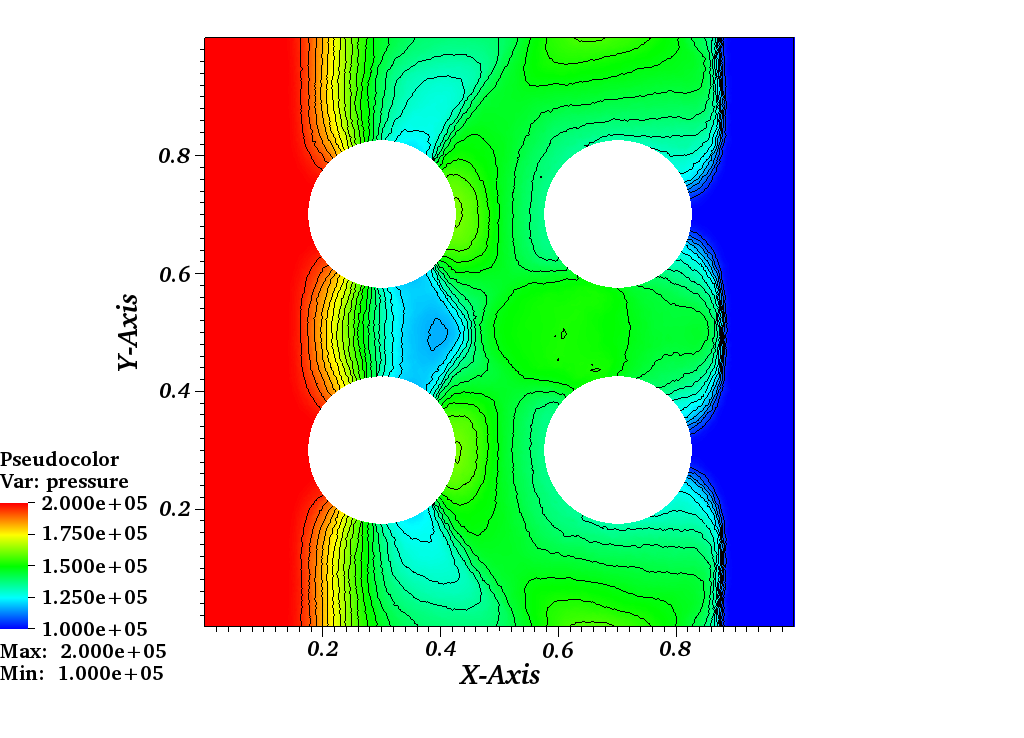}\label{2D_CBS_pressure}}
\caption{2D liquid-vapor water mixture with mass transfer at $t=0.8$~ms. Residual Distribution method based on a non-conservative formulation of the system \& MOOD. Density $\rm \left[kg/m^3\right]$ and  pressure $\rm [Pa]$ representation on an unstructured grid with $N=79512$ elements.
}\label{CBS_2D_test}
\end{figure}

\begin{figure}[H]
\centering
\includegraphics[width=0.49\textwidth]{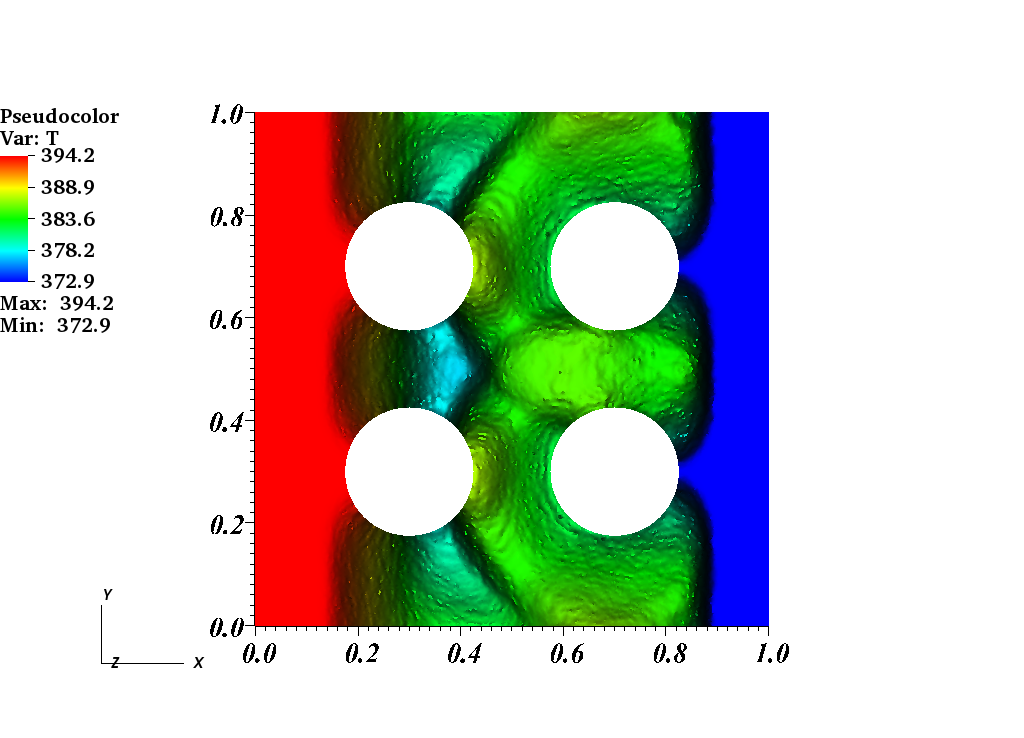}
\includegraphics[width=0.49\textwidth]{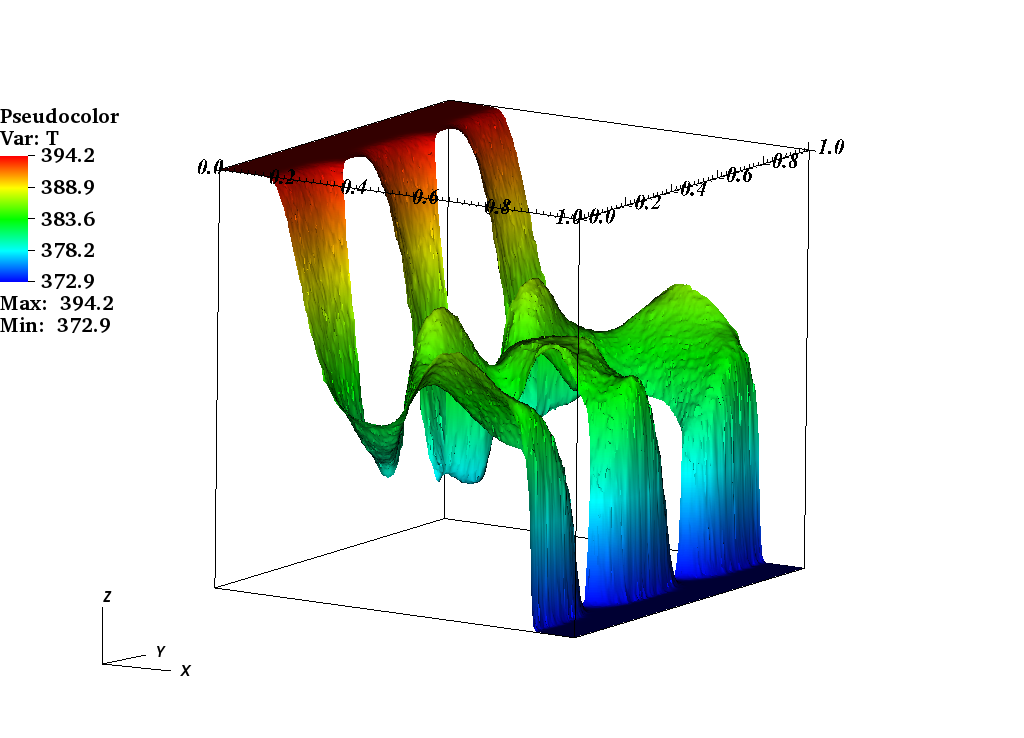}
\caption{2D liquid-vapor water mixture with mass transfer at $t=0.8$~ms. RD approach. Temperature $\rm [K]$ field from helicopter view (left) and in a 3D representation over the domain (right) for $N=79512$ elements.
}\label{2D_CBS_T_elev}
\end{figure}
\begin{figure}[H]
\centering
\includegraphics[width=0.6\textwidth]{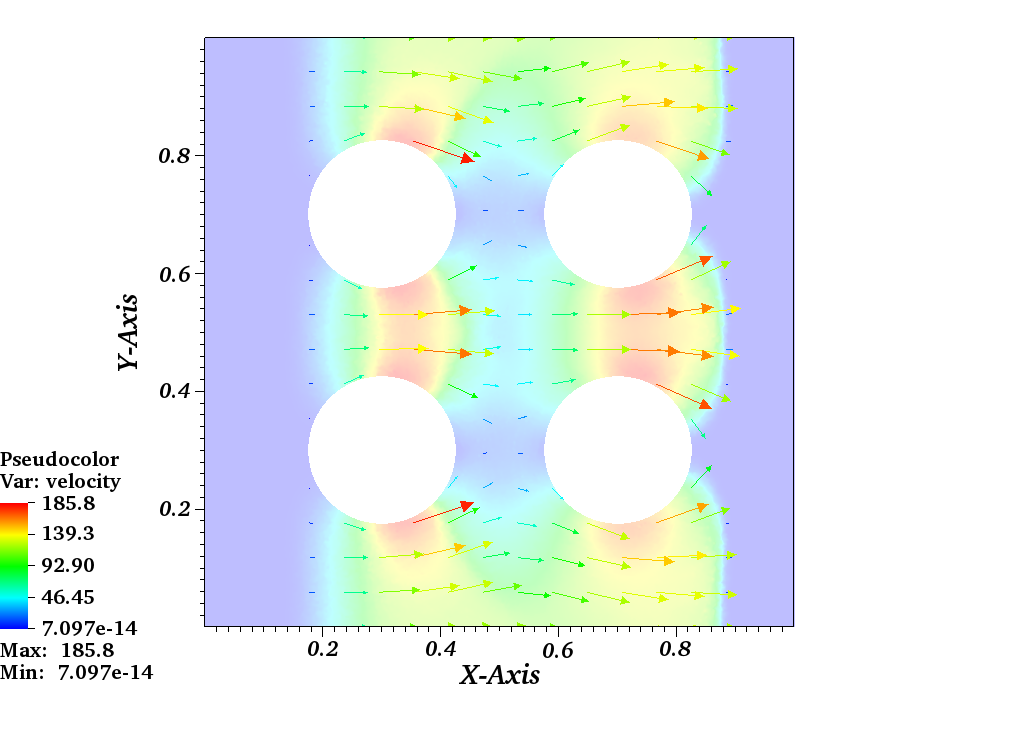}
\caption{2D liquid-vapor water mixture with mass transfer at $t=0.8$~ms. RD approach. Velocity field on an unstructured grid with $N=79512$ elements.
}\label{2D_CBS_velocity}
\end{figure}
\begin{figure}[H]
\centering
\includegraphics[width=0.5\textwidth]{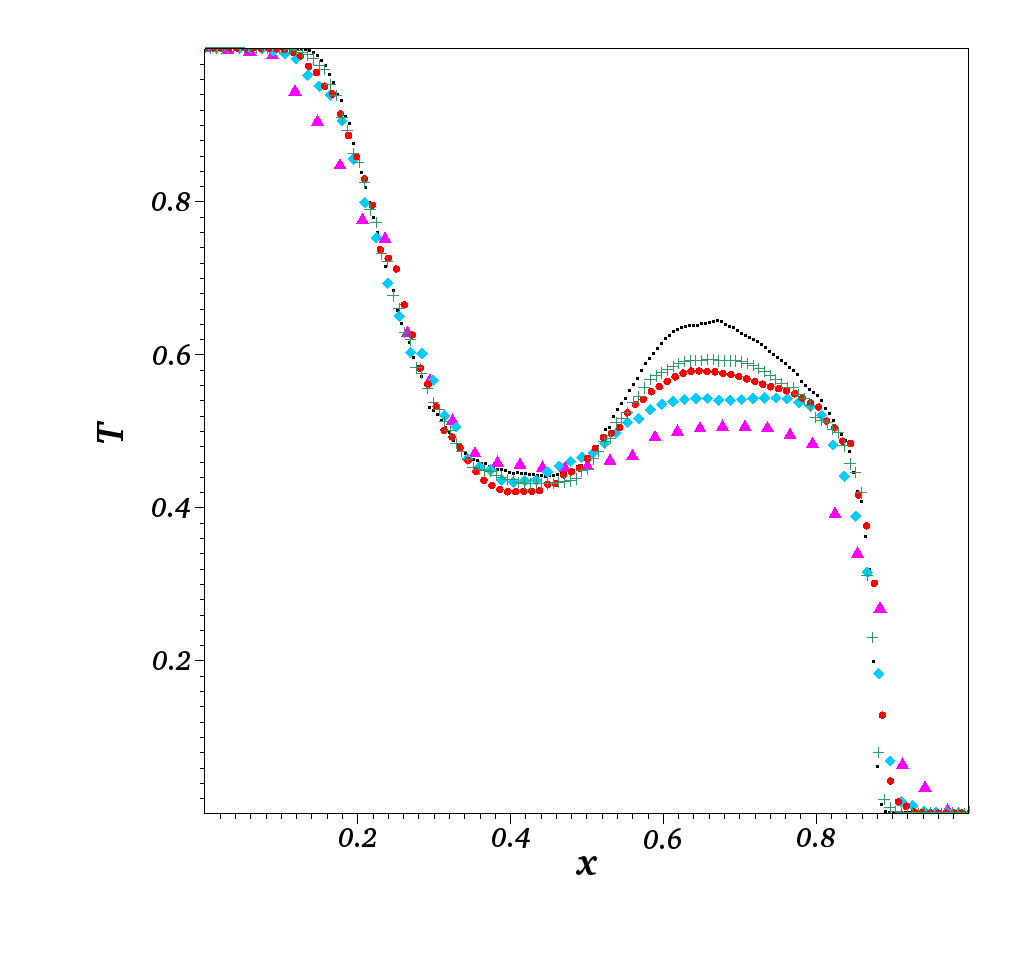}
\caption{2D liquid-vapor water mixture with mass transfer at $t=0.8$~ms. RD approach. Scatter plot of the scaled temperature (normalized axis to a cube) for different unstructured grids: magenta/tetrahedron $N=2696$ elements, blue/octahedron $N=10780$ elements, red/icosahedron $N=20306$ elements, green/axis $N=40198$ elements, black/points $N=79512$ elements.
}\label{2D_CBS_T}
\end{figure}

As CP is designed to rely on structured grids, we will hereafter only consider the RD approximation, which is designed to work on both structured and unstructured grids. Figure \ref{CBS_2D_test} shows the density in Fig. \ref{2D_CBS_density} and the pressure in Fig. \ref{2D_CBS_pressure}  at $t=0.8$~ms along with contour lines. In particular, one can observe a drop of the density, as well as of the pressure for $0.3\,{\rm m}<x<0.5\,{\rm m}$ caused by the presence of the cylinders, which is not observed in the 1D related benchmark of Figure \ref{cbs_mass}. In Figure \ref{2D_CBS_velocity} the velocity, along with its vector field representation is shown, showing an overall good symmetry of the velocity profile. As for the temperature, we show in Figure \ref{2D_CBS_T_elev} a helicopter view on the left, along with a 3D display on the right that shows the temperature elevation on the z-axis along the x- and y-axis of the domain. To evaluate the benchmark in terms of mesh refinement, Figure \ref{2D_CBS_T} shows the scatter plot of the scaled temperature, which considers a normalized axis to a cube, for five different unstructured grids. In particular, for the coarsest mesh with $N=2696$ elements we displayed the normalized temperature via  magenta tetrahedrons, the second mesh with $N=10780$ elements  via blue octahedrons, the intermediate one with $N=20306$ elements via red icosahedrons, the fourth mesh with $N=40198$ elements through green axis and, finally, the finest mesh, used also for the previous mentioned representations, with $N=79512$ elements via black points. Overall one can see, that the finer the mesh is considered, the steeper the area between 0.6~m and 0.8~m gets, while in the region around 0.2~m and 0.4~m the profile gets sharper, but does not show a neat difference. 

\section{Conclusions}\label{sec:conclu}

In this paper, we have presented a second-order Finite Element based Residual Distribution scheme with an ``a posteriori limiting'' for a four-equation multiphase flow model written in a non-conservative form, with source terms modelling mass transfer.
We have tested this novel methodology on several severe and diverse benchmark problems and we have cross-validated the approximated solutions with a HLLC solver written for the classical conservative form of the system. Overall, we have observed a good agreement between the two numerical methods and have assessed the robustness and convergence on several problems in multi-dimensions  with or without phase transition.
The forthcoming investigations within this context will include an extension to a higher than second-order approximation through a Finite Element based Residual Distribution method that follows the work presented in~\cite{AbgrallHO2018}.

\begin{acknowledgements}
P. B. has been funded by the SNSF project grant $\# 200021\_153604$. 
\end{acknowledgements}

%
%



\bibliographystyle{spmpsci}      
\bibliography{biblio}
%
%

\end{document}